\documentclass[pre,twocolumn,showpcs,floatfix,amsmath,amssymb,titlepage]{revtex4}
\usepackage{pslatex,graphicx,dcolumn,bm,natbib}
\usepackage{bm}        
\usepackage{mathrsfs} 
\usepackage{color}

\newcommand{\be}{\begin{equation}}
\newcommand{\ee}{\end{equation}}
\newcommand{\bea}{\begin{eqnarray}}
\newcommand{\eea}{\end{eqnarray}}
\newcommand{\eqnarr}{\begin{eqnarray}}
\newcommand{\eqnend}{\end{eqnarray}}
\newcommand{\abs}[1]{\left| #1 \right|}

\newcommand{\st}{\hat \sigma}

\begin{document}
\title {Shear-induced Rigidity of Frictional Particles:  Analysis of Emergent Order in Stress-Space}
	\author{
	Sumantra Sarkar$^1$, Dapeng Bi$^{2,3}$, Jie Zhang$^4$, Jie Ren$^5$, R. P. Behringer$^5$, and Bulbul Chakraborty$^1$\\	 
	$^1$ Martin Fisher School of Physics, Brandeis University, Waltham, MA 02453, USA\\     
	$^2$Department of Physics, Syracuse University,
	Syracuse, NY 13224, USA\\
	$^3$Center for Studies in Physics and Biology Rockefeller University, New York, NY 10065, USA \\
	$^4$Institute of Natural Sciences and Department of Physics, Shanghai Jiao Tong University, Shanghai 200240 China\\
	$^5$Department of Physics, Duke University, Durham, NC, USA}

\begin{abstract}
{ Solids are distinguished from fluids by their ability to resist shear. In equilibrium systems, the resistance to shear is associated with the emergence of broken translational symmetry as exhibited by a non-uniform density pattern that is persistent, which in turn results from minimizing the free energy.  In this work, we  focus on a  class of systems where this paradigm is challenged. We show that shear-driven jamming in dry granular materials is a collective process controlled by the constraints of mechanical equilibrium. We argue that  these constraints can lead to a persistent pattern in a dual space that encodes the statistics of contact forces and the topology of the contact network.  The shear-jamming transition is marked by the appearance of this persistent pattern.  We investigate the structure and behavior of  patterns both in real space {and} the dual space as the system evolves through the rigidity transition {for} a range of packing fractions {and} in two different shear protocols.   We show that,  in the protocol that creates homogeneous jammed states without  shear bands, measures of shear jamming  do not depend on strain and packing fraction independently but obey a scaling form with a packing-fraction dependent characteristic strain that goes to zero at the isotropic jamming point, $\phi_J$.  We demonstrate that it is possible to define a protocol-independent order parameter in this dual space, which provides a quantitative measure of the rigidity of shear-jammed states.}


\end{abstract}

\maketitle


\newcommand{\IncludeFigureFTIntro}{\begin{figure}[htbp]
\includegraphics[width=\linewidth]{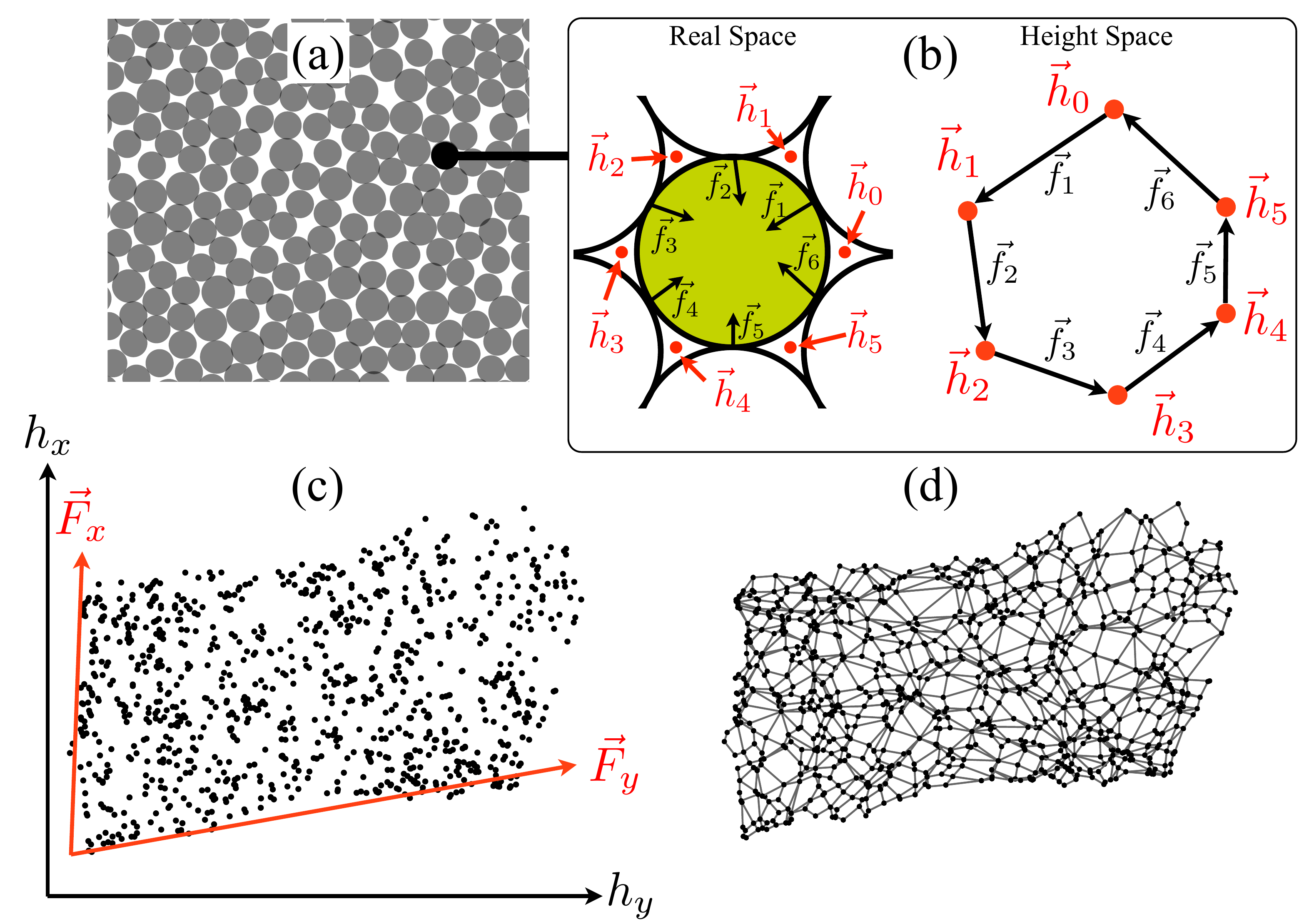}
\caption
{(Color online)
Force tiles and $\lbrace h_{i} \rbrace$ of a typical  experimental  SJ state 
(a) The real-space configuration of grains. 
(b) Height field defined on the voids (red points) around a single grain. 
Starting from an arbitrary origin, and going around the grain in a counterclockwise direction, the height, $\vec{h}_\nu$,  is incremented by  the contact force, $\vec{f}_i$,  separating two voids. 
The vectors $\vec{f}_i$ form a closed polygon when adjacent forces are arranged head-to-tail. The vertices of this polygon (force-tile) are given by the values of the heights $\vec{h}_\nu$. 
(c) The height vertices. 
(d) The force tile network corresponding to  the height vertices shown in (c). $ \vec{F_x} $ and $ \vec{F_y} $ indicates the extent of the applied external stress. The compressive direction (larger force) is chosen according to~\cite{BiNature2011}.  This figure has been reproduced from \cite{Sarkar2013Origin}.
}
\label{forcetile}
\end{figure}}


\newcommand{\IncludeFigureTorus}{

\begin{figure}[htbp]
\begin{center}
\includegraphics[width=0.45\textwidth]{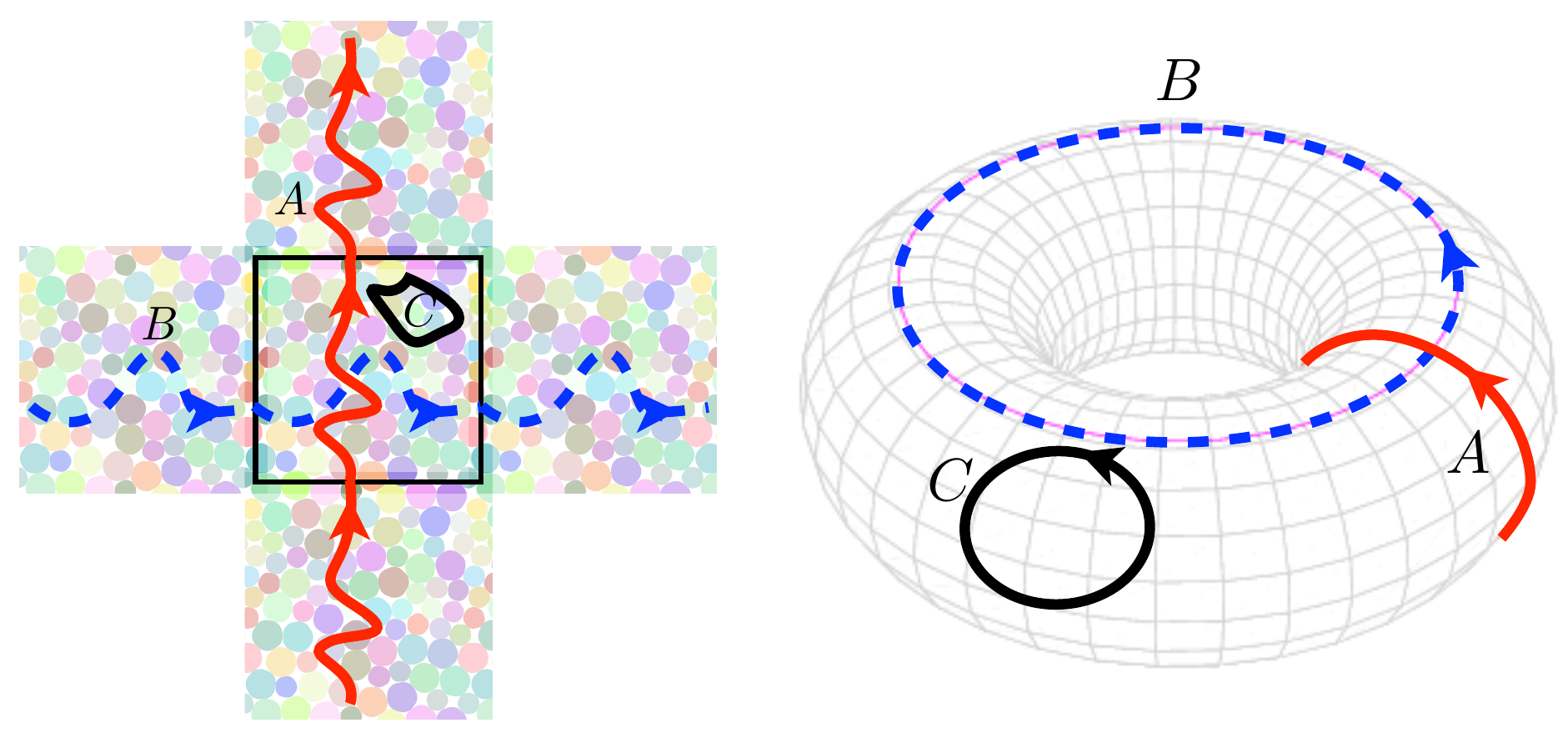}
\caption
{(Color online)
Illustration of Invariants: Left: The physical system, a 2-D
granular solid of size $ L \times L $, is outlined by the black box, also shown
are its images under PBC. The lines A \& B represent the two distinct
classes of non-contractible loops in the system and C represents a trivial
loop. Right: Representation of the system on the surface of a torus. Loops
A \& B are non-contractible and correspond to the same labeling as in
the left panel. To change $ \vec F_x (\vec F_y) $, a change has to be made on the non-contractible
loop B (A).
}
\label{fig:torus}
\end{center}
\end{figure}

}


\newcommand{\IncludeFigureConvexityCriteria}{
\begin{figure}[htbp]
\begin{center}
\includegraphics[width=0.45\textwidth]{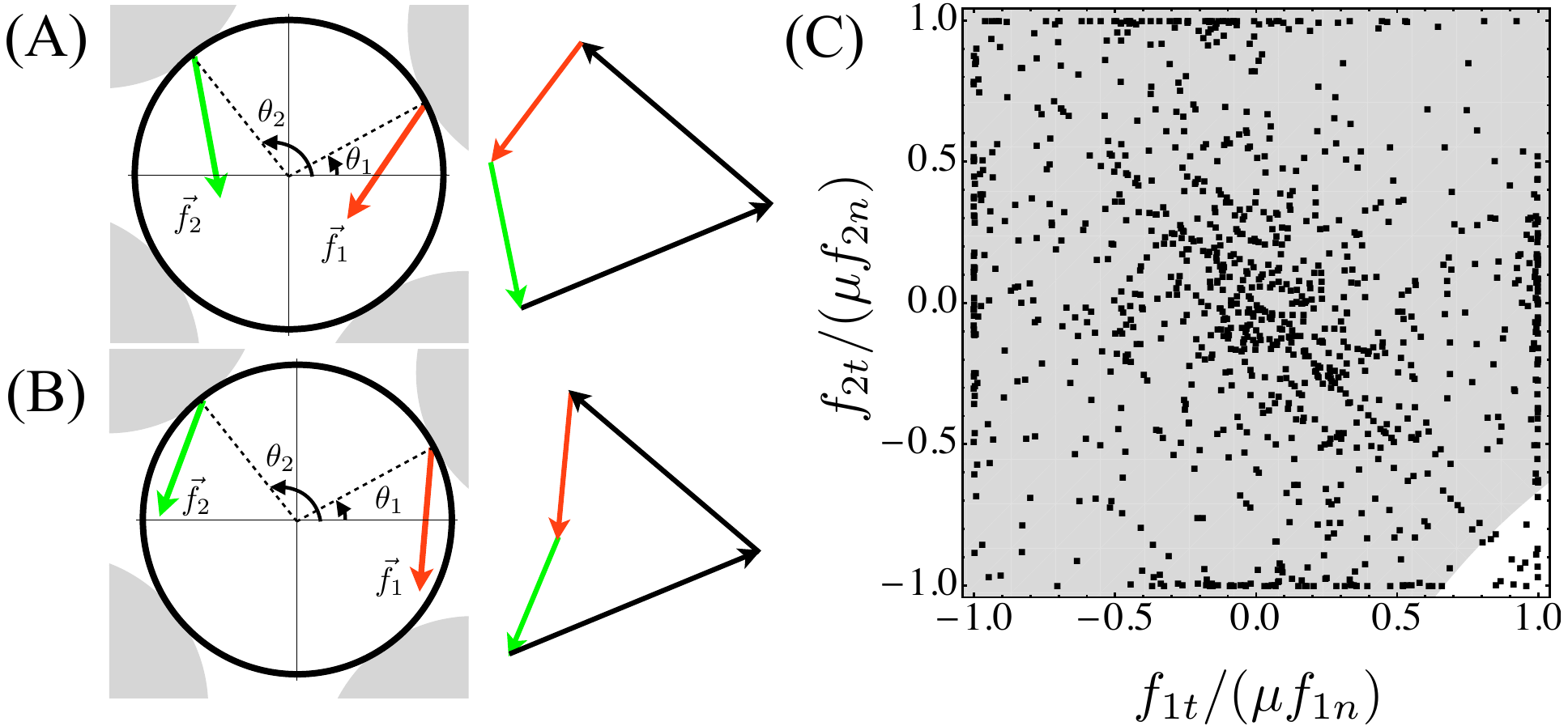}
\caption
{(Color online)
\textbf{Convexity criterion for force tiles:} 
(A) A typical case where two consecutive forces form a part of a convex polygon. (B) A rare case where two consecutive forces form a part of a concave polygon.  Convexity is determined by  the sign of  $\vec{f}_1 \times \vec{f}_2$, which depends upon the angular separation of adjacent  contacts ($\theta_2-\theta_1$), and the magnitude of the tangential components of these forces. (C) Convexity map for $\theta_2-\theta_1 \ge \pi/3$  (Eq.~\ref{convexity_cond}) and $\mu=0.7$ (packing fraction $ \phi = 0.805 $ and strain $ \gamma = 15\% $). Grey region denotes convex and white concave. Experimental data from a typical SJ state is also shown (points).  {This figure has been reproduced from \cite{Sarkar2013Origin}.}
}
\label{convexity}
\end{center}
\end{figure}
}


\newcommand{\IncludeFigureApparatus}{
\begin{figure}
\centering 
\includegraphics[width=\linewidth]{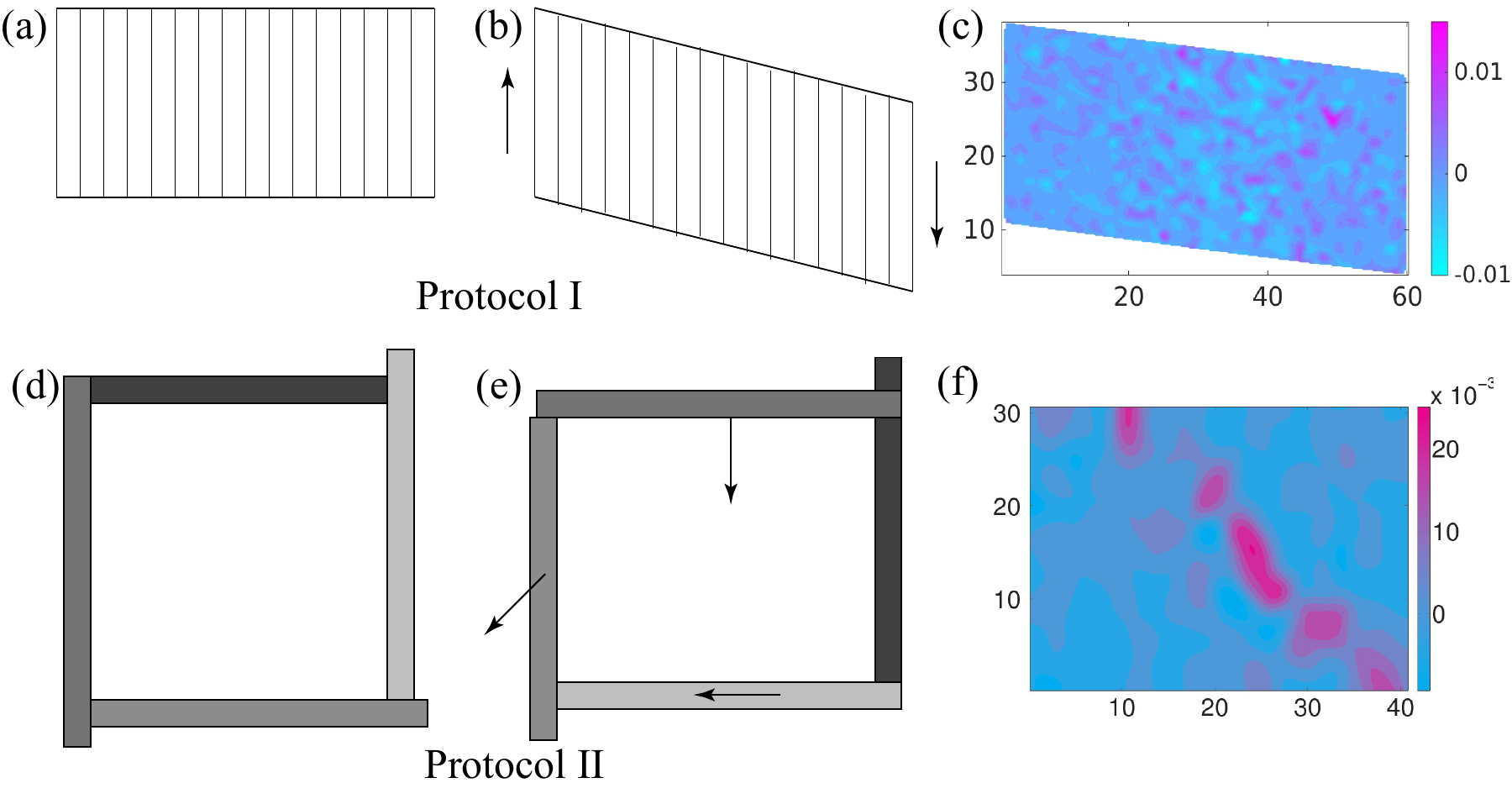}
\caption{(Color online) Schematics of shear strain devices. The first device
is sketched in parts (a) and (b). Photoelastic particles rest on a
surface that consists of smooth powder-coated slats cut from
transparent Plexiglas. During an experiment, the walls and slats
deform at constant area, and the whole system undergoes simple shear,
as indicated by the arrows of part (b). (c) Incremental strain field ($ \delta \epsilon_{xx} $) between $ \gamma = 13.5 \% $ and $ 13.77\% $ from the first device shows small homogeneous fluctuation.  The second device is sketched
in parts (d-e). With this apparatus, we can independently control the
spacing between opposing walls to achieve arbitray strains. For the
present experiments, we only consider strains where the are of the
enclosed rectangle is constant, corresponding to pure shear strains.(f) { Incremental strain field ($ \delta \epsilon_{xx} $) between $ \gamma = 9.24 \% $ and $ 9.57\% $, from the second device shows macroscopic strain inhomogeneity.} }
\label{fig:apparatus}
\end{figure}

}


\newcommand{\IncludeFigureFTAlgo}{
\begin{figure}
\centering
\includegraphics[width=\linewidth]{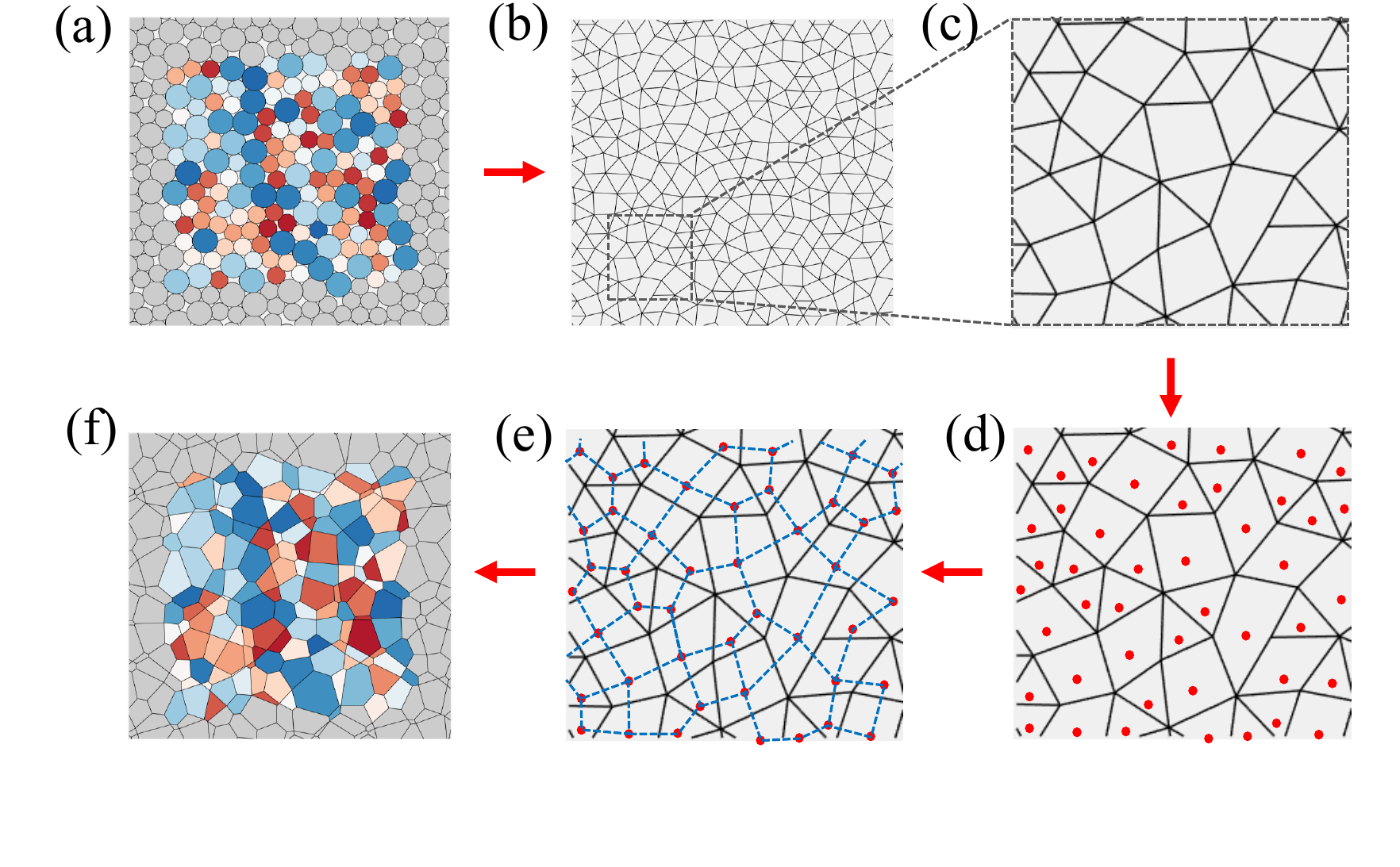}
\caption{(Color online) A schematic of the force tiling algorithm. (a) A typical grain configuration. The colors are used to tag a grain and the corresponding force tile, otherwise they have no physical significance. (b) The real space contact network. (c) A portion of the real space contact network. (d) The faces (a point inside every face is marked by a red dot) of this portion of the real space contact network as obtained from the MCB algorithm. (e) The dual graph (blue dashed line) topology obtained from the MCB. (f) The force tiling for this configuration. The color of the force tile matches the color of the corresponding grain in (a). }
\label{fig:FT_Algo}
\end{figure}

}


\newcommand{\IncludeFigureGrainPosition}{

\begin{figure*} [htbp]
\centering
\includegraphics[width=\textwidth]{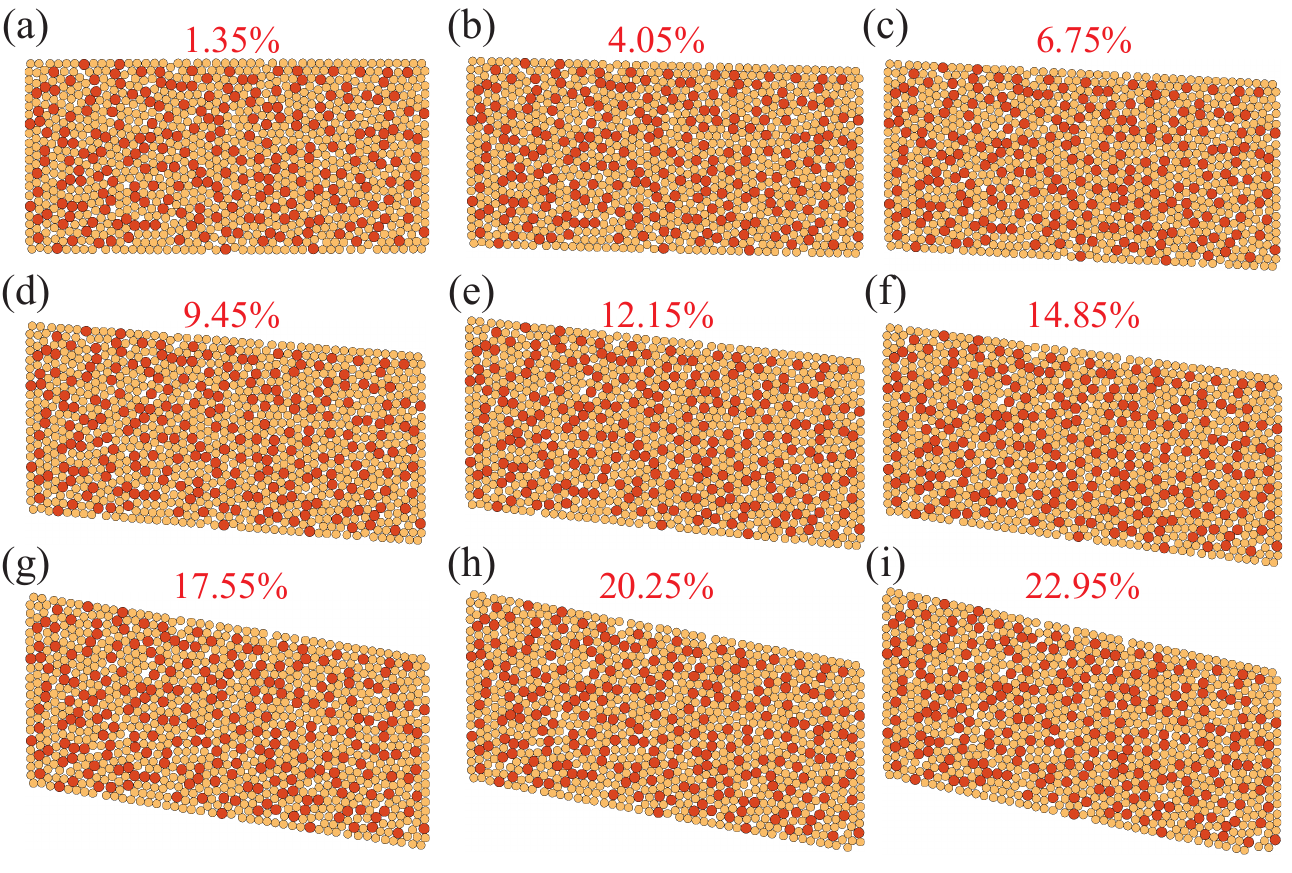}
\caption{(Color online) {\bf Grain Positions:} The evolution of the grain positions in a shear jamming experiment at $\phi = 0.8163$. Bidisperse grains (dark orange: larger and light orange: smaller grains) were used to avoid crystallization. As the system is sheared the positions of the grains remain virtually unchanged {( evinced by the overlap matrix of the real-space position in Fig.~\ref{fig:OL_Comp} )} apart from the global affine deformation, characterized by the global strain marked on every snapshot.    }
\label{fig:grain_position}
\end{figure*}
}


\newcommand{\IncludeFigureForceTiling}{\begin{figure*} [htbp]
\centering
\includegraphics[width=\textwidth]{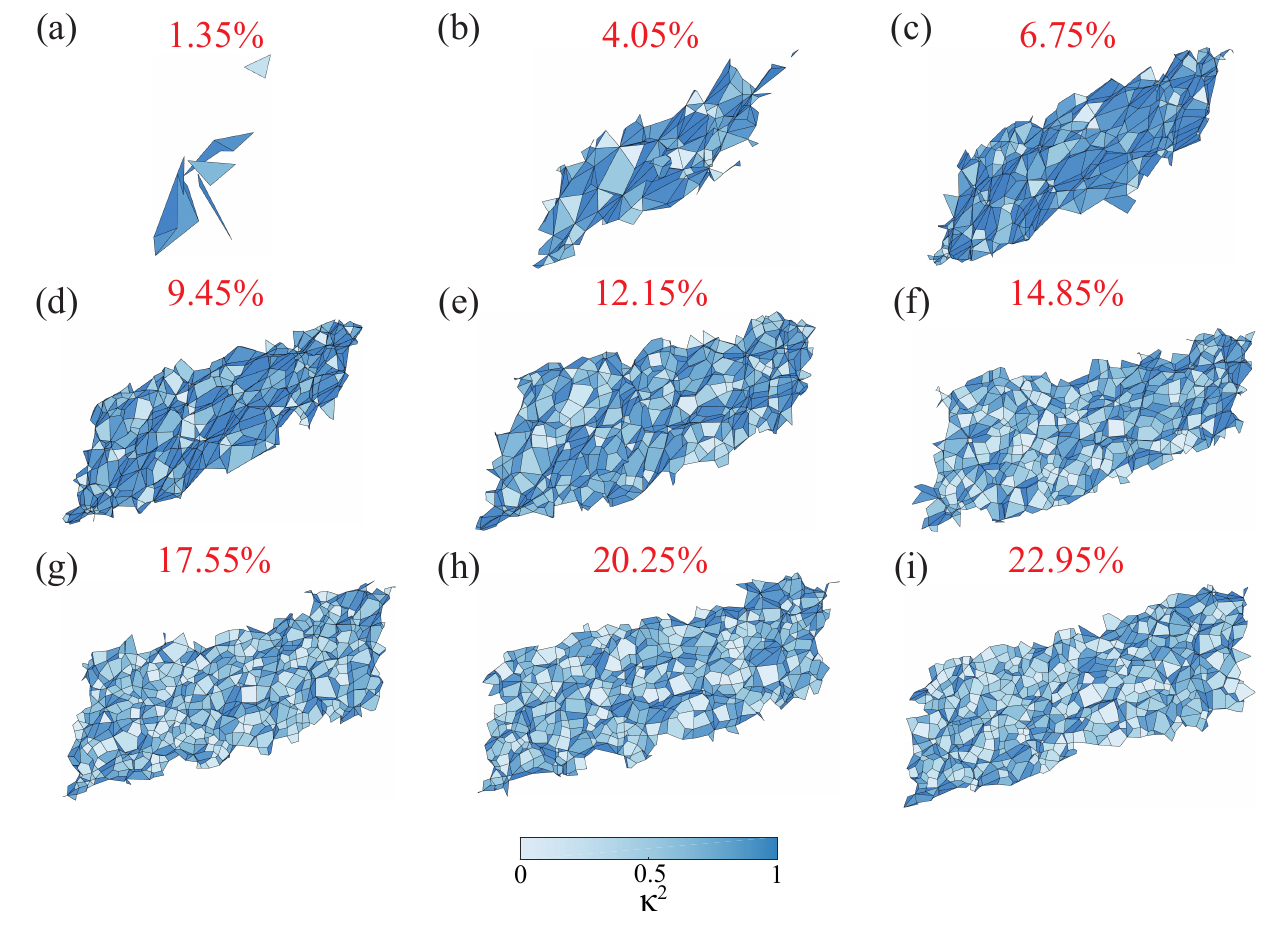}
\caption{(Color online) {\bf Force Tiling:} The evolution of the shape of the force tiles in a shear jamming experiment at $\phi = 0.8163$. For the sake of clarity, we have not shown the evolution of the size of the force tiles, which increases with increasing shear strain (marked on the snapshot). Each force tile is colored according to its asphericity (see appendix), $ \kappa^2 $. In the unjammed state (a), the force tiling is very small (due to small forces) and formless. In the fragile state (b,c,d) the force tiling are very anisotropic (as characterized by high asphericity of individual tiles) and begins as a quasi one dimensional structure(b), which evolves towards a well defined two dimensional shape as the shear jamming approaches (d). In the jammed states (e-i), the tiling has a well-defined shape which remain preserved even when a large amount of strain is applied. Also, individual tiles become more isotropic.}
\label{fig:FT_SJ}
\end{figure*}}


\newcommand{\IncludeFigureOLComparison}{

\begin{figure} [htbp]
\centering
\includegraphics[width=0.5\textwidth]{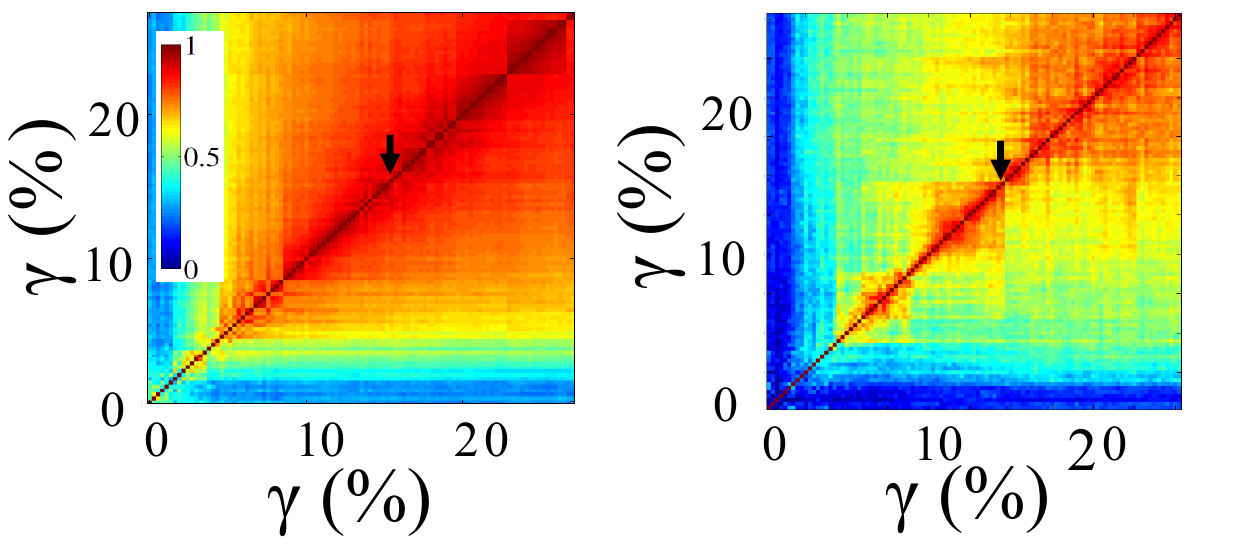}
\caption{(Color online) The overlap matrix of the RSNs (left) and the FTNs (right) for protocol I at $\phi = 0.8163$. The black arrow marks the onset of the shear jamming transition, as detected from the saturation of $f_{NR}$\cite{BiNature2011}.   }
\label{fig:OL_Comp}
\end{figure}

} 


\newcommand{\IncludeFigureEpsilonStar}{
\begin{figure*}[tbph]
\centering
\includegraphics[width=0.8\linewidth]{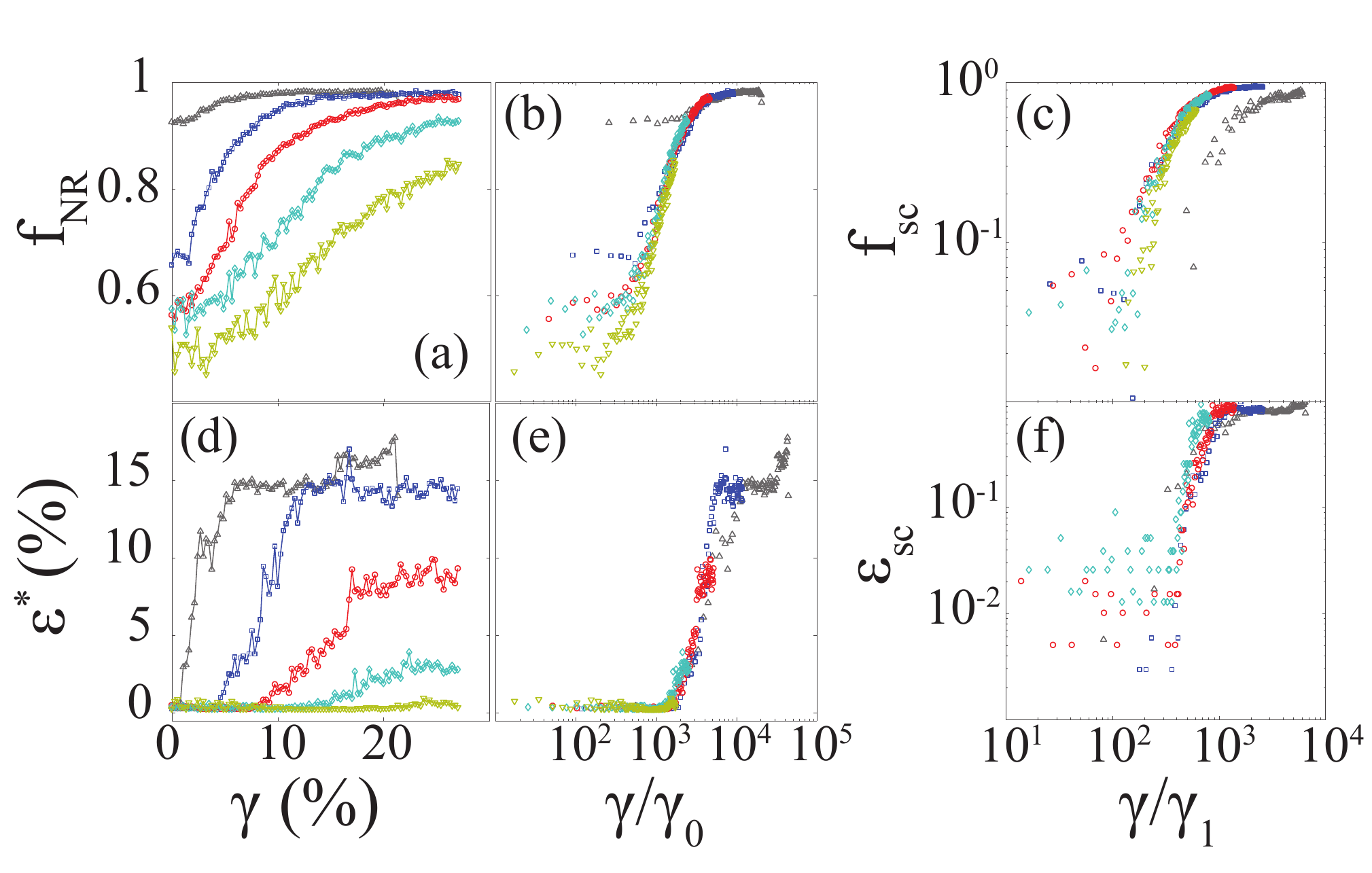}
\caption{(Color online) Dependence of $ f_{NR} $ (a, b, c) and $\epsilon^*$ (d, e, f) on strain for packing fractions: 0.8269 (gray,$\bigtriangleup$), 0.8163 (blue, $\Box$), 0.8036 (red, $\bigcirc$), 0.7863 (turqoise, $\Diamond$) and 0.7728 (yellowish-green, $\bigtriangledown$) in protocol I. $ \epsilon^* $ measures the {\it range} of $\gamma$ with 50\% or more overlap as a function of the strain $ \gamma $ itself. (a) and (d) shows variation of $ f_{NR} $ and $ \epsilon^* $ with strain $ \gamma $. (b) and (e) Scaling of $ f_{NR} $ and $ \epsilon^* $ with $ \gamma $;  $ \gamma_0 =  (1-\phi/\phi_J)^{1.6} $. (c) and (f) Scaling of $ f_{sc} $ and $\epsilon_{sc}$ (see text for definitions) with $ \gamma $; $ \gamma_1  = (1-\phi/\phi_J)^{1.2}$. }
\label{fig:epsilon}
\end{figure*}

}


\newcommand{\IncludeFigureJRComparison}{
\begin{figure}[tbph]
\centering
\includegraphics[width=0.5\textwidth]{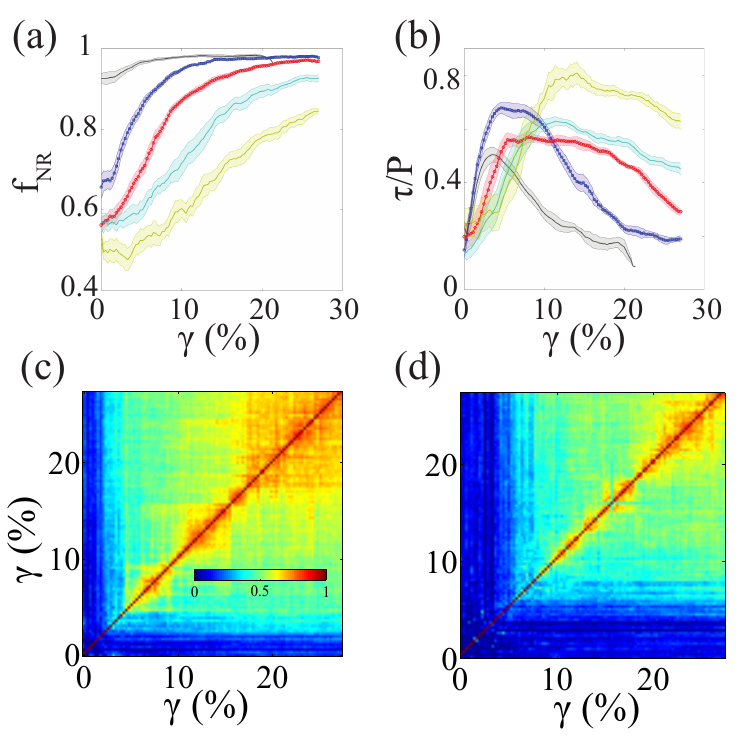}
\caption{(Color online) Shear jamming experiments with protocol I. (a) and (b) shows the evolution of, respectively, the $ f_{NR} $ (reproduced from Fig. \ref{fig:epsilon} for ease of comparison)  and the stress anisotropy $ \tau/P $ as a function of strain $ \gamma $, for five different packing fractions:0.8269 (gray), 0.8163 (blue, $\Box$), 0.8036 (red, $\bigcirc$), 0.7863 (turqoise) and 0.7728 (yellowish-green).  The curves are arranged in the order of increasing packing fraction from bottom to top.
The solid line is the average from five different runs at each packing fraction, whereas the shaded area shows the standard deviation of the mean. The stress anisotropy peaks at the jamming transition concomitant with the saturation of the $ f_{NR} $. The strain at which this transition happens increases with decreasing packing fraction.  In the bottom panel, we compare the force space overlap matrix for packing fractions 0.8163  (c) and 0.8036 (d) . The overlap matrices show that the onset of shear rigidity (regions with high overlap) occurs at larger strains for smaller packing fractions.}
\label{fig:JRComparison}
\end{figure}
}


\newcommand{\IncludeFigureJRJZFT}{

\begin{figure}[tbph]
\centering
\includegraphics[width=\linewidth]{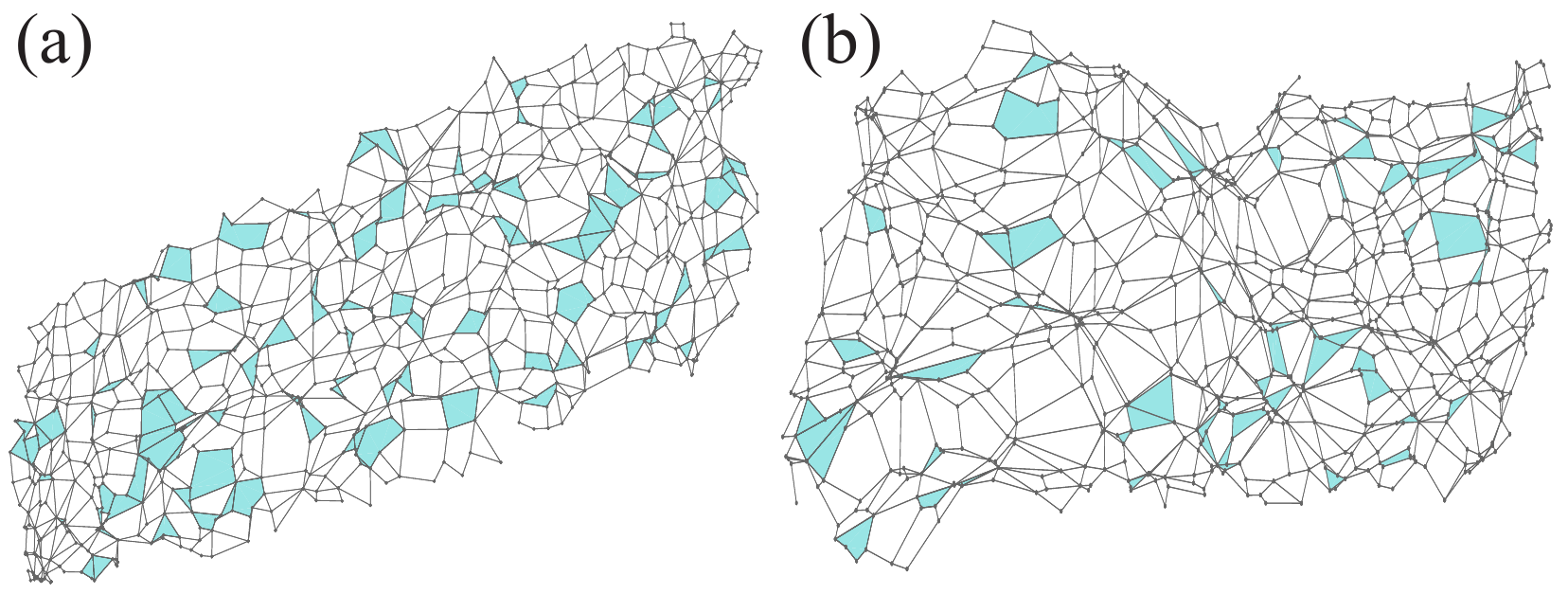}
\caption{(Color online) Force tiling from a shear jammed configuration for protocol I (a), and protocol II (b); the shaded tiles are the non-convex polygons. Due to the inhomogeneity of applied strain, the forces in protocol II are also inhomogeneous and are larger at one boundary compared to the other. This is reflected in the trapezoidal shape of the force tiling in protocol II, where the tiles are larger on the right side compared to the left. }
\label{fig:JRJZ_FT}
\end{figure}
}


\newcommand{\IncludeFigureAsphericity}{

\begin{figure}[tbph]
\centering
\includegraphics[width=\linewidth]{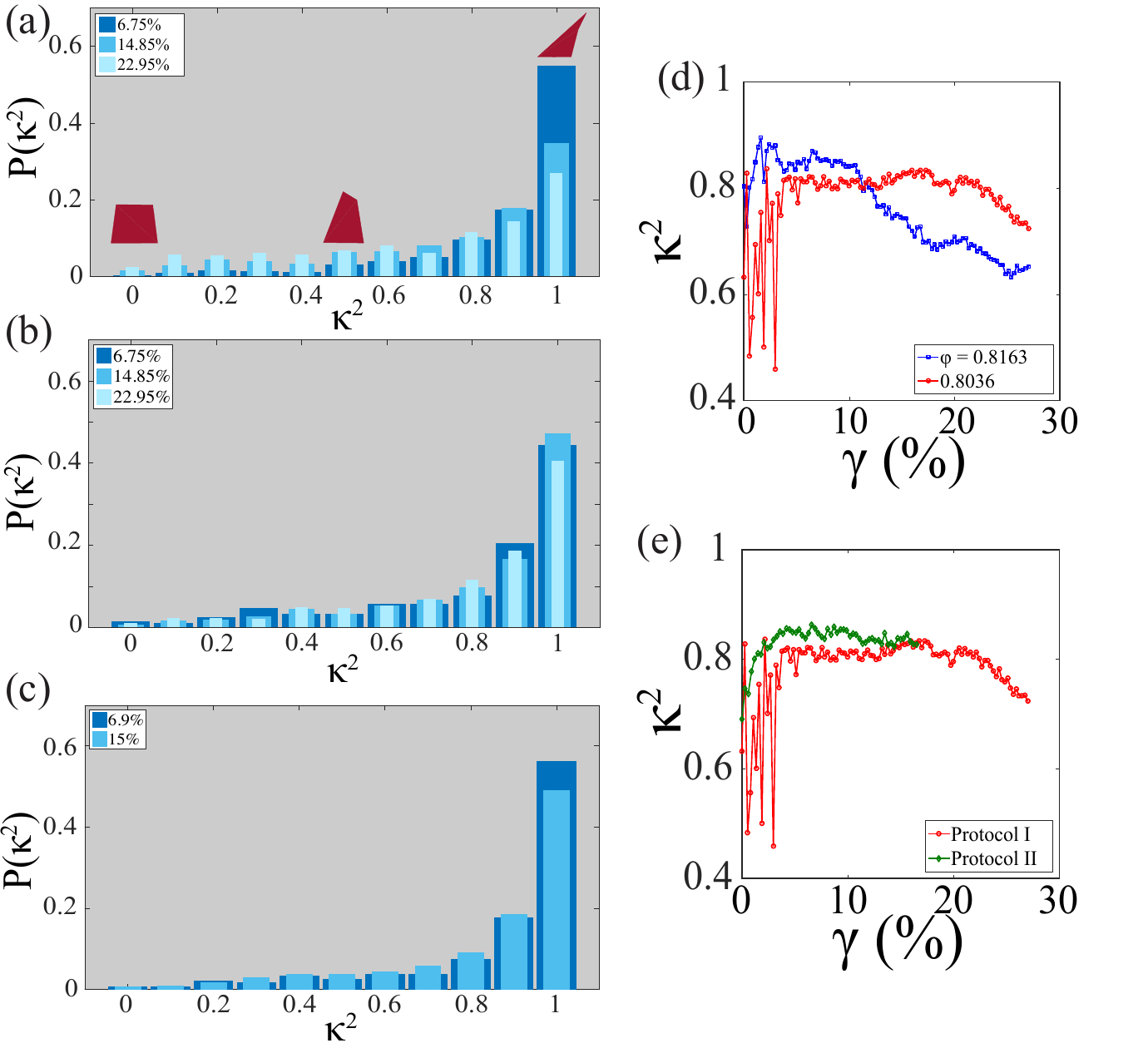}
\caption{{(Color online) Evolution of asphericity of tiles ( see Appendix ) during the shear jamming proces. The distribution of the asphericity of the tiles at different shear strains (legend) during the shear jamming transition with (a) protocol I; $ \phi = 0.8163 $ (b) protocol I; $ \phi = 0.8036 $, and (c) protocol II; $ \phi =0.8036 $. Few representative tiles are also shown. The asphericity peaks at 1 for all strains, but with increasing strain, the peak strength decreases and the distribution becomes broader, indicating proliferation of isotropic tiles. (d) Mean asphericity as a function of strain at two different packing fractions (legend) in protocol I. (e) Mean asphericity as a function of strain for two different protocols at $ \phi = 0.8036 $. Protocol II has higher mean asphericity compared to the protocol I.} }
\label{fig:Asphericity}
\end{figure}

}


\newcommand{\IncludeFigureJRJZOne}{

\begin{figure}[tbph]
\centering
\includegraphics[width=\linewidth]{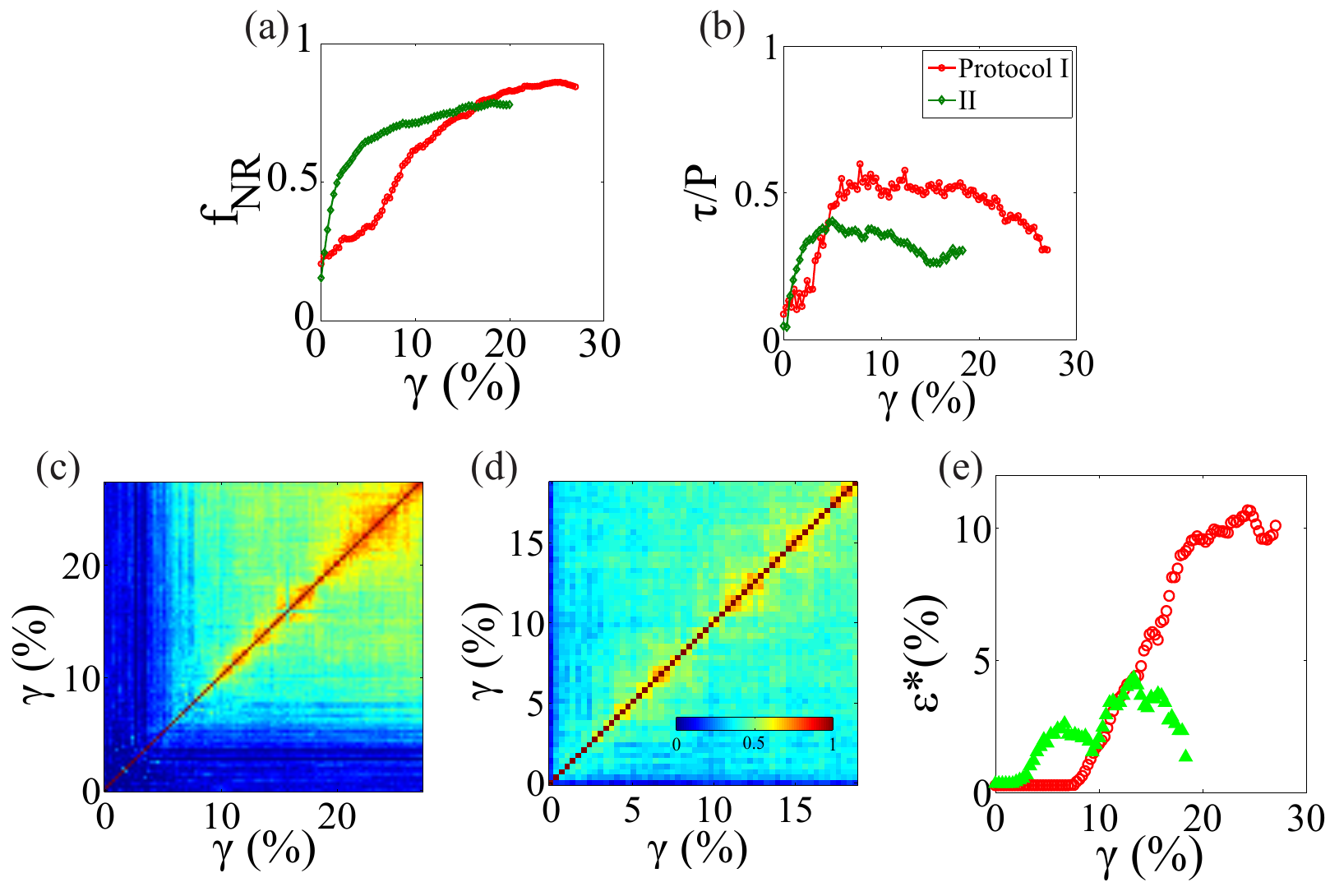}
\caption{(Color online) \textbf{Comparison of protocols:} (a) $ f_{NR} $ in the protocol II saturates at a lower strain compared to the protocol I. The saturation value is smaller also. (b) The stress anisotropy peaks at a lower strain and the peak value of the stress anisotropy is lower in \textbf{protocol II} than the protocol I. The overlap matrix in protocol II (d) is distinctively different than protocol I (c). Even though the overlap matrix for protocol II reaches a overlap value of $\sim 0.5$ pretty quickly, unlike protocol I it never attains really high ($ > 0.8 $) overlap, which suggests lack of persistent order even in the jammed state. (e) $ \epsilon^* $ for protocol I (red circle) and protocol II (green triangle) $ \phi = 0.8036 $. }
\label{fig:JRJZ_1}
\end{figure}
}


\newcommand{\IncludeFigureJRJZThree}{

\begin{figure}[tbph]
\centering
\includegraphics[width=\linewidth]{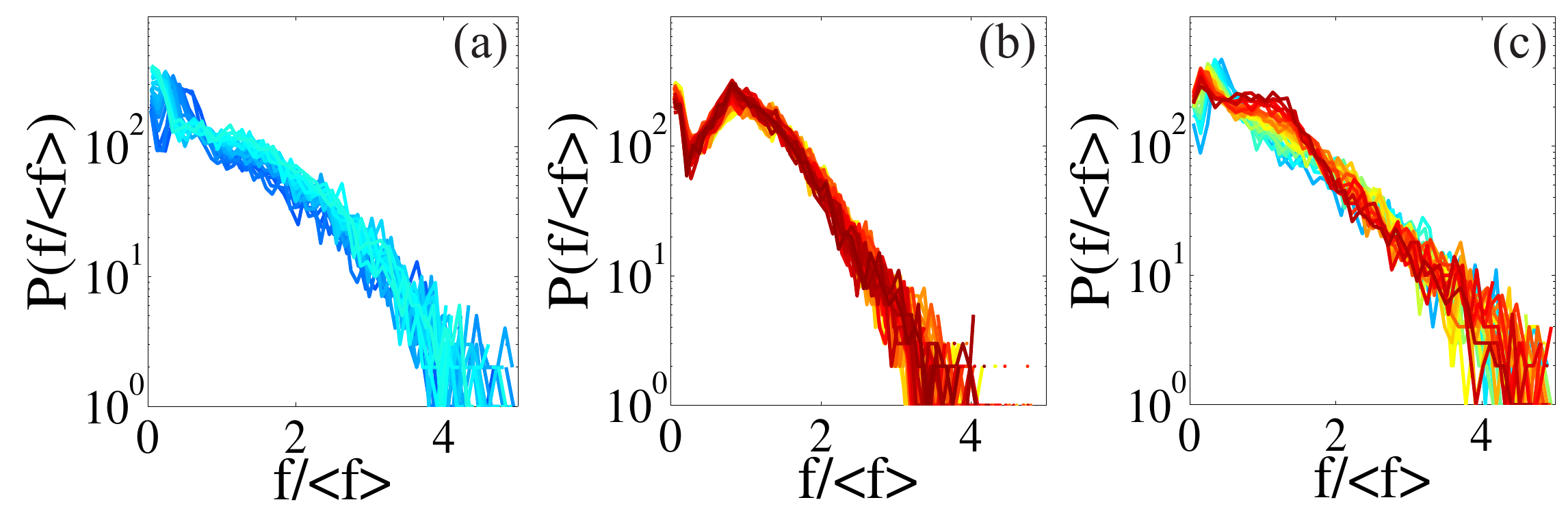}
\caption{(Color online) Force distributions at $ \phi = 0.8036 $. (a) Fragile and (b) jammed packings generated by protocol I. (c)  jammed packing generated by protocol II. The shear strain increases from blue to dark red. Dark red states are jammed states.}
\label{fig:ForceDistro}
\end{figure}
}


\newcommand{\IncludeFigurePercolation}{
\begin{figure}[htb]
\center
\includegraphics[width=0.85\linewidth]{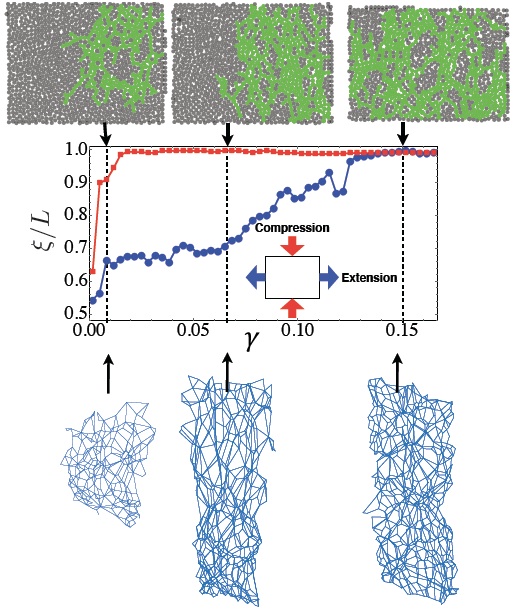}
\caption
{
(Color online) Top: Plots of the cluster size to box dimension ratios ($\xi/L$) vs. strain for a shear jamming process via protocol II at $\phi=0.805$. Pure shear strain (compression in the $y$-direction and extension $x$ while total area is fixed) is applied to a initially unjammed system. Red line corresponds to the $y-$size of the largest contiguous cluster with contact force $f>\langle f \rangle $ in the system. Blue line corresponds to the $x-$size of the largest contiguous cluster with contact force $f>f_{avg}$ in the system. Snapshots of clusters are shown for $\gamma = 0.01$ (unjammed), $\gamma = 0.066$ (fragile) and $\gamma = 0.15$ (shear jammed).  Bottom: Force tilings for same set of configurations.   The force tilings exhibit changes in global shape and local structure.   
}
\label{fig:clusters}
\end{figure}

}


\newcommand{\IncludeFigureJRJZTwo}{
\begin{figure}[tbph]
\centering
\includegraphics[width=\linewidth]{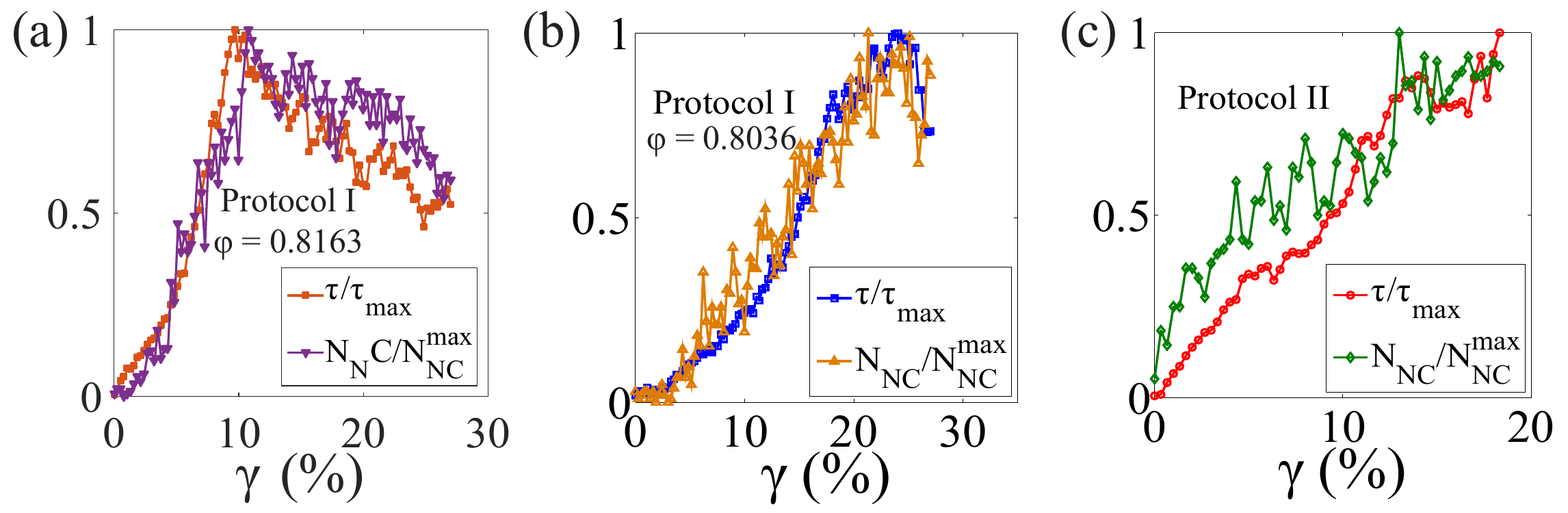}
\caption{(Color online) The number of non-convex polygons, normalized by the maximum number reached during the strain history,  and $ \tau $ as a function of shear strain for (a) protocol I at $ \phi = 0.8163 $ (b) protocol I at $ \phi = 0.8036 $, and (c) protocol II at $ \phi = 0.8036 $. In (a) and (b), both $ \tau $ and $ N_{NC} $ peaks at the strain value at which $ \epsilon^* $ reaches its maximum value. } 
\label{fig:JRJZ_NNC}
\end{figure}

}


\newcommand{\IncludeFigureStressGeometryRelation}{

\begin{figure}
\centering
\includegraphics[width=\linewidth]{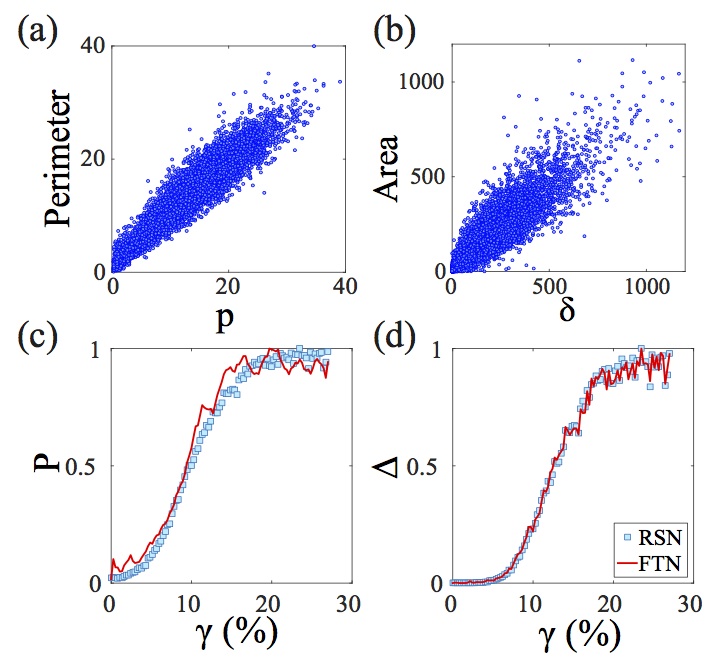}
\caption{(Color online) Scatter plot of (a) pressure, $ p $, of individual grains and the perimeter of the corresponding force tile and (b) determinant of the stress tensor of individual grains, $ \delta $, and the area of the corresponding force tile. (c) The ``global" pressure (sum of the pressures of all the grains), $ P $, and the perimeter of the boundary of the force tiling, and (d) determinant, $ \Delta $, of the force moment tensor and the area of the force tiling as a function of strain. All variables in (c) and (d) are scaled by their maximum value for easy comparison. The data is from the experiment done at $ \phi = 0.8163 $ using protocol I. }
\label{fig:Stress_Geometry_Corr}
\end{figure} 

}



\section{Introduction}
 Solid packings of dry grains are amorphous structures  that are created through external driving at zero temperature.       In the last couple of decades, much progress has been made in analyzing the problem of the glass transition~\cite{karmakar2009growing,parisi2010mean,kurchan2011glasses,biroli2011glass},  and the nature of rigidity of  disordered solids.    In these systems, either density or temperature {changes induce} a transition from a fluid to an amorphous solid.    The amorphous solid acquires a shear-rigidity because of a persistent pattern in the positions of  the particles~\cite{Alexander}:   a pattern that is difficult to quantify~\cite{Kurchan_Levine} but is known to exist.    In spite of their structural similarities to amorphous solids at finite temperatures,   the question of what imparts rigidity to granular solids poses additional challenges because of the absence of any cohesive interactions between the grains and because of the absence of thermal fluctuations.   { In this paper, we address the question of rigidity of dry grains by analyzing the persistence of patterns in both positions and contact forces.}
  
Jamming of frictionless soft grains,  which has been extensively studied over the last two decades~\cite{van2010jamming,corey_simul} occurs deep in the glass phase and is characterized by the onset of mechanical equilibrium.  This phenomenon can be described within the concept of isostaticity~\cite{corey_simul,Wyart:2005kb} and the jamming transition occurs at some protocol dependent density at which the structure becomes isostatic~\cite{Chaudhuri:2010fk}. 
``Jamming'' has become synonymous with this rigidity  transition of frictionless grains.   An alternative picture of  jamming was, however, proposed in the late '90's~\cite{Cates1998},  prompted by observations in  non-Brownian  suspensions of colloidal particles.   This scenario can be described as shear-induced solidification:    a fluid to solid transition where the solidity emerges solely as a result of applied stresses. {One expects that in such a transition  there is an organization in the space of forces that drives or stabilizes the positional patterns.}   Moreover, the driving  mechanism for collective organization of the grains has to be the constraints of mechanical stability subject to the {\it globally} imposed stresses:  a scenario that is  different from {the} density or temperature-driven glass transition.  A crucial feature that distinguishes the shear-jamming process from the density-driven jamming process is its inherently anisotropic character, and the presence of  a bath  of  ``spectator''  grains that do not bear any forces.  The shear-jammed (SJ) structure forms within this bath as more and more grains become incorporated into the force-bearing network in response to the externally imposed shear.  In contrast,  isotropic, density-driven jamming occurs in the absence of a bath:  except for a vanishingly small fraction in the thermodynamic limit,  all grains are part of the force-bearing network.   
    
Analysis of recent experiments on shear-jamming in dry grains~\cite{BiNature2011,Ren2013Thesis,ren2013reynolds} and discontinuous shear thickening in dense, non-Brownian suspensions~\cite{Seto2013,mari2014shear,Brown2010,Brown2013} have found remarkable similarities between experimental observations and the theoretical picture of stress transmission in shear-jammed solids that was presented in the 90's.   In this paper, we  present  a theoretical framework that quantifies the phenomenon of shear-induced jamming.   In particular, we construct an order parameter that distinguishes these solids from fluids.   The theoretical framework is applied to experimental studies of shear jamming in dry grains.   Part of  this work has been published earlier in a concise form~\cite{Sarkar2013Origin}.

This paper is organized as follows.    In section II, we develop a description  of granular solids in a space that is ``reciprocal'' to position space,  in a sense to be defined below.   In section III, we describe different experimental protocols and present an algorithm for generating the reciprocal space representation from a knowledge of contact forces and positions of grains.   In section IV,  we propose an order parameter that can be identified with shear-jammed states, which indicates that shear-jamming is associated with a broken-symmetry in this reciprocal space. In this section, we also analyze experimental data using the theoretical perspective developed in the earlier sections, and in section V, we present our conclusions and ideas for future work.  


\section{Theoretical framework}

{  A collection of dry grains interacting via purely repulsive contact interactions can be viewed in some respects as a system of hard particles.   Solidification of hard particles is entropic in origin and requires the presence of thermal fluctuations.  The rigidity of these solids arises from an effective ``cohesion''  caused  by entropic forces. In contrast to such Brownian systems of hard particles, the lack of thermal fluctuations in dry grains imply  that  there are no intrinsic mechanisms to heal broken contacts and generate cohesive interactions.  An alternative mechanism that can lead to   a solid-like response to shearing  is the collective organization of grains arising from the constraints of mechanical equilibrium.   

 {A dry granular packing has to satisfy four {types} of constraints that arise from the requirements of the mechanical equilibrium at zero temperature.} The  constraints of force and torque balance have to be satisfied for every grain. Since the contacts are frictional, the Coulomb criterion of static equilibrium has to be satisfied.  This introduces an additional constraint, {$ \left| f_t\right| \le \mu \left| f_n \right|$}, where $ \mu $ is the coefficient of friction and $ f_{t(n)}$ is the tangential (normal) component of the contact force. The interaction between  dry grains is purely repulsive, hence the normal force  has to be positive, which is an additional inequality constraint. As discussed below, the equality constraint of force balance and the inequality constraint of positive normal forces can be incorporated by resorting to a geometric representation, dual to the real-space geometry. The torque balance condition and the Coulomb inequality  manifest themselves by affecting the patterns in this dual space.     }

\subsubsection{Mechanical equilibrium and height fields} 

The continuum analog of the force and the torque balance conditions can be expressed in terms of  the stress tensor $ \st $: $ \vec \nabla\cdot\st = \vec{F}, \st = \st^T $. 
In a two dimensional packing with only boundary loading ($ \vec{F} = 0 $),   $ \vec \nabla\cdot\st = 0 $
can be enforced through the introduction of a vector field of gauge potentials referred to as height vectors, $ \vec h(\vec r) $~\cite{Henkes,ball-blumenfeld,degiuli2012continuum,degiuli2013granular,Bi2015AnnRev}: $\st  =  \vec  \nabla\times \vec h$. Here $ \vec  \nabla\times \equiv \hat\epsilon\cdot\vec \nabla $, and $ \hat{ \mathbf{\epsilon}} $ is the 2D Levi-Civita tensor; the $ ``\cdot" $ refers to matrix multiplication~\cite{degiuli2012continuum}. Hence, 
\begin{eqnarray}
\st &=& \vec \nabla\times \vec h \\
    &=& \hat{ \mathbf{\epsilon}}\cdot\vec \nabla \vec h\\ 
    &=& \begin{pmatrix} \partial_yh_x & \partial_yh_y \\-\partial_xh_x & -\partial_xh_y \end{pmatrix}\label{eq:curl_h}.
\end{eqnarray}
 Since  torque balance requires that the stress tensor is symmetric, $\vec \nabla\cdot\vec{h} = 0 $. 
 

This two dimensional continuum description can be derived from an equivalent discrete formulation at the grain level~\cite{Henkes,ball-blumenfeld,degiuli2012continuum,degiuli2013granular,Bi2015AnnRev}. There, the heights are uniquely defined on the dual space of the contact network or the voids surrounding the grains. A geometric representation which omits the real space geometry, but retains the topology of  the contact network, accurately represents the structure in the space of height vectors. The vectors representing the heights appear as vertices of a network in which, starting from an arbitrary origin (gauge freedom), force vectors are laid down to generate the edges connecting these vertices. Since we enforce force balance for each and every grain, and two touching grains share an equal and opposite force (Newton's third law), the heights form the vertices of a tiling of the plane by polygons. This network of polygons is equivalent to the Maxwell-Cremona tiling~\cite{maxwell1890scientific} or force tiling~\cite{Tighe,Sarkar2013Origin},  where each polygonal face represents a grain.  


\IncludeFigureFTIntro  


\subsubsection{Structure of height space}
Under periodic boundary condition, any 2D system can be mapped to the surface of a torus~\cite{Bi2012Thesis}. The integral of $ \st $ over a topologically trivial (contractible) loop (Fig.~\ref{fig:torus}, curve $ C $) vanishes identically, because $ \vec  \nabla\cdot\st =0 $. However, for non-contractible {loops} (Fig.~\ref{fig:torus}, curve $ A $ or $ B $), which spans the system, the integral is nonzero. These two integrals: 

\IncludeFigureTorus
\begin{eqnarray}
\vec F_x = \int_0^{L_y} dy \begin{pmatrix}
\sigma_{11}(x,y) \\
\sigma_{12}(x,y)
\end{pmatrix};\quad
\vec F_y = \int_0^{L_x} dx\begin{pmatrix}
\sigma_{21}(x,y) \\
\sigma_{22}(x,y) 
\end{pmatrix} 
\end{eqnarray}
are topological invariants of the system. $ \left(L_x, L_y\right) $ are the size of the system along the corresponding directions. Physically, $\left(\vec F_x,\vec F_y\right) $ amount to the total load along the two directions, and are related to the force-moment tensor.
\begin{eqnarray}
\hat \Sigma = \sum_{\langle ij  \rangle} \vec r_{ij}\otimes \vec f_{ij} = \begin{pmatrix}
L_x & 0 \\
0 & L_y
\end{pmatrix} \times \begin{pmatrix}
\vec F_x \cdot \hat x & \vec F_x \cdot \hat y\\
\vec F_y \cdot \hat x & \vec F_y \cdot \hat y
\end{pmatrix}
\end{eqnarray}
Here, the sum defining the force-moment tensor is over all contacts, $ \vec r_{ij} $ is the contact vector from the center of grain $ i $ to the inter-particle contact between grains $ i $ and $ j $, and $ \vec f_{ij} $ is the force associated with that particular contact. The pressure, $ P $, of the packing is given by $ (\lambda_1 + \lambda_2)/2 $, and the anisotropy of the stresses is given by $ \tau = |\lambda_1 - \lambda_2|/2 $, where $ \lambda_{1,2} $ are the eigenvalues of the stress tensor of the entire packing, $ \st_{global} = \hat \Sigma /N $; $ N $ is the total number of grains. The stress anisotropy of the global stress tensor is $ \tau/P $. Since $  \vec \nabla\times\vec h = \st $, $\left(\vec F_x,\vec F_y\right) $ represent the net change of height across the sample ~\footnote{This can be seen in the following way. Let's consider $ \vec F_x = \int_0^L\left[\sigma_{11}(x,y), \sigma_{12}(x,y)\right]^T dy$. Now, from equation \ref{eq:curl_h}, we know that $ \sigma_{11} = \partial_yh_x $ and $ \sigma_{12} = \partial_yh_y $. Integrating we get $ \vec F_x  = \vec h(x,L) - \vec h(x,0)$, which is nothing but the difference of the height fields across the sample in the y direction. }. Consequently, the force tiling in the height space is confined within a parallelogram formed from $\left(\vec F_x,\vec F_y\right) $; only a non-zero applied stress leads to a finite structure in the height space. Since these two vectors are topological invariants of the system, one can construct a statistical ensemble of all force tiles confined within this parallelogram as the analog of a microcanonical ensemble in equilibrium statistical mechanics.  In this ensemble, energy is replaced by the vectors  $\left(\vec F_x,\vec F_y\right) $.   The force network ensemble (FNE) approach to granular elasticity~\cite{snoeijer2004force,tighe2008entropy} is an example of such an ensemble, as is the generalized Edwards ensemble~\cite{Bi2015AnnRev,BC-softmatter}. This description can be generalized to a finite system.   The shape of the confining region is, however,  no longer a parallelogram. An average parallelogram can be defined via the columns of the force-tensor $ \hat \Sigma $, as shown in Fig.~\ref{forcetile}.  

\subsubsection{Positivity, Coulomb criterion, and convexity}

Any force balanced configuration leads to a height pattern, $ \rho(\vec h) = \sum_i \delta\left(\vec h - \vec h_i\right) $. If such a height pattern does not change under small, continuous deformation of the boundary, we will define such a structure to have persistent order in height space. It is analogous to how one may define rigidity for an elastic solid, where the rigidity measures the persistence of the density field of the atoms of the material when the system is strained. Since changes to the boundaries of the force tiling is equivalent to changing the boundary load on the sample, a granular assembly, created at a given $\left(\vec F_x,\vec F_y\right) $, will collectively resist shear deformation, if it has persistent order. The question that we ask is whether the condition of mechanical equilibrium can lead to persistent order, and under what conditions. 

Since forces can be arbitrarily small, the heights are continuous variables. A set of heights confined within the parallelogram bounded by $\left(\vec F_x,\vec F_y\right) $ represents a force-balanced configuration. Thus, in the absence of any other constraints, the heights should not show any correlation or broken symmetry. The torque balance condition, the positivity of the normal forces, and the Coulomb criterion, however, provide additional constraints on the geometrical structure of the height space.  

{The positivity of the normal forces guarantees that the height field increases monotonically.  Therefore, there is only a single height origin and therefore, a single sheet of tiles.   In the presence of attractive and repulsive interactions the polygons would be organized on multiple sheets since there would be multiple points in the packing where the height would go to zero.}

For a granular solid composed of frictionless, grains of convex shapes (circles or ellipses, for example) with only normal forces between grains, a rotation of 90$^\circ$ converts all forces to  tangential. A convex polygon that exactly inscribes the grain can be constructed by simply elongating the rotated force vectors. This polygon is related to the force tile by a conformal transformation. Hence, in the absence of frictional forces,  all force tiles are convex. This convexity constraint is equivalent to the torque balance condition for frictionless grains, and of course,  the Coulomb condition is always satisfied.

It is possible to have non-convex polygons as force tiles when frictional forces exist. Two consecutive forces around a {disk} can either form a convex vertex (Fig.~\ref{convexity} A) or a concave one depending on how frictional they are and the angular distance $\theta_2-\theta_1$ between the contacts. Decomposing each force into a tangential and a normal part, the condition for convexity can be easily obtained:
\begin{equation}
1+\frac{f_{1t} f_{2t}}{f_{1n} f_{2n}} +\left(\frac{f_{1t}}{ f_{1n}}-\frac{f_{2t}}{ f_{2n}}\right) \cot(\theta_2-\theta_1) \ge 0, 
\label{convexity_cond}
\end{equation}
where the tangential force and the normal force obey the Coulomb criterion for a given static friction coefficient $\mu$: $\frac{f_{1t}}{f_{1n}}, \frac{f_{2t}}{f_{2n}} \le \mu$. The angular distance between two contacts $\theta_2-\theta_1$ is constrained by geometry.   In a mono-dispersed packing of just-touching disks, for example,  $\theta_2-\theta_1$ cannot be smaller than $\pi/3$. Using this as a lower bound, Eq.~\ref{convexity_cond} gives the range of values of tangential forces for which convexity is possible for a given $\mu$. A straightforward calculation based on  Eq.~\ref{convexity_cond} shows that for any $\mu<1/\sqrt{3}\simeq 0.58$, the convexity condition is never violated. For  $\mu=0.7$, which is the static friction coefficient of the particles studied in the experiments~\cite{BiNature2011,granmat_bob}, it is possible to have non-convex polygons when the two consecutive forces $\vec{f}_1$ and $\vec{f}_2$ are simultaneously fully mobilized contacts ($ |f_t| = \mu f_n $), or are close to being fully mobilized.  In Fig.~\ref{convexity}(C), we perform this analysis on a typical experimental shear-jammed state ~\cite{zhang2010jamming,BiNature2011} created  at $\phi = 0.805$  under a pure-shear strain of 15\%, and for grains with a friction coefficient of $\mu=0.7$. The rescaled tangential forces for all pairs of consecutive contacts are represented by a scatter plot. The convexity criterion given by Eq.~\eqref{convexity_cond} is represented by the shaded region in Fig.~\ref{convexity}(C).  While a few contact pairs form concave edges, they are rare occurrences and we deduce that, statistically, in the shear jammed states,  the force tiles are convex for typical physical values of $\mu$.


\IncludeFigureConvexityCriteria 



{ The Coulomb criterion $|f_{t}| \le \mu f_{n}$ is  the most difficult to implement within the tiling representation~\cite{Degiuli}.   The previous discussion indicates that these constraints can be effectively captured as a convexity constraint on the force tiles  of stable packings since non-convex tiles occur only when the tangential forces are close to the failure threshold.}
If unconstrained in height space, the ensemble of all possible point patterns formed by the vertices are trivially expected to have a liquid-like order or $\langle\rho(\vec h)\rangle=const$.
With the requirements of convexity and the strict edge-matching constraints of tiling, the vertices of a tile cannot come arbitrarily close to each other. This requirement constrains the possible point patterns formed by the vertices of the tiles to a much smaller subset of configurations, hence giving rise to the possibility of broken translational symmetry in height space or  $\langle\rho(\vec h)\rangle \ne const$.
 The constraints act as effective springs that tie the vertices to their average positions. If these springs constrain the position of every vertex in the tile to a region that is small compared to the average force (length of a link), then we expect to see correlations and broken translational invariance in height space.  The strengths of the effective springs are not predetermined but emerge as a consequence of the local constraints and the global constraints through $\vec F_x$, $\vec F_y$, or $\hat \Sigma$.   

Based on the above points, we argue  that broken translational symmetry and persistent order emerges in height space as the number of vertices is increased  through the creation of force-bearing contacts between grains as a set of grains is stressed.   In the remainder of this paper, we construct and analyze height patterns of experimentally generated shear-jammed states, and show that rigidity is concurrent with appearance of persistent patterns of heights and occurs at a critical value of the fraction of force-bearing grains~\cite{BiNature2011}.   { The appearance of the persistent  pattern  is also accompanied by a decrease in the number of non-convex polygons in homogeneous shear-jammed states.}


\section{Experimental methods and analysis of experimental data}

We apply the above-mentioned theoretical framework to experimental systems that exhibit shear jamming, and explore the jamming dynamics from dual space representations. For this purpose, we examine two different sets of shear experiments on two dimensional systems of photo-elastic particles. These experiments not only reliably demonstrate shear-jamming behaviors according to force and stress analyses, as presented elsewhere~\cite{zhang2010jamming,BiNature2011}, but also provide full data sets of microscopic location and contact forces of each particle, which can be readily used in the dual-space construction.

For protocol I the shear apparatus, Fig.~\ref{fig:apparatus} (a-b), acts to deform a rectangular region into a parallelogram (simple shear), while preserving the total system area. The basic scenario of the experimental approach has been presented in Ren et al.~\cite{ren2013reynolds}. For protocol II, Fig.~\ref{fig:apparatus} (d-e), the shear apparatus starts from a square shape, then contracts along one dimension and elongates along the other, while keeping the total area constant (pure shear). The base of this apparatus is a smooth Plexiglas plate that does not move, thus the shear is purely boundary-driven. Many details of the experiments have been published elsewhere~\cite{Ren2013Thesis,BiNature2011,granmat_jie,granmat,ren2013reynolds}.  For completeness, we provide additional relevant details on the experiments and protocols in Appendix I.

\subsection{Experimental protocols}
These strain-controlled experiments involve two different types of experimental
apparatus, which apply shear strain $ \gamma $ to the sample.  
The  simple shear apparatus (protocol I) has a special
arrangement of slats on the bottom, as sketched in
Fig.~\ref{fig:apparatus} (a-b), which shows an overhead schematic of
the apparatus. In this system, shear is applied at the walls, and also
uniformly through the base. This means that particles which are
rattlers or experience very weak forces from surround grains are not
left behind during the initial phases of shear. As a result, the
system exhibits locally coarse-grained shear strain that corresponds
to the global affine shear,  with small, spatially homogeneous, fluctuations~\cite{Ren2013Thesis}, as shown in Fig.~\ref{fig:apparatus}(c) . Shear bands or other macroscopic inhomogeneities do
not develop in this system.

An overhead schematic of the second shear apparatus (protocol II) is sketched in Fig.~\ref{fig:apparatus}
(d-e). The boundaries of this system are controlled so as to produce
pure shear, consisting of compression in one direction and dilation
in the orthogonal direction.  The particles rest on a base consisting of a smooth
Plexiglas sheet and are confined by the boundaries, whose positions
are controlled by a pair of stepper motors (not shown). The upper
boundary is fixed in the frame of the base. The three other walls move
as indicated in part (d). That is, in order for the system to evolve
from (c) to (d), the three lower boundaries move in the indicated
directions relative to the base and top wall. The side walls are
maintained in a rectangular geometry by {guides}. This apparatus allows
a deformation of the boundaries in a continuous range of rectangular
geometries. However, for the experiments described here, the area of
the interior, which contains the particles, is held fixed.  Thus,  the
strains correspond to pure shear. In this device, the strain is
applied strictly at the boundaries, as is typical of most granular
strain devices. Consequently, there is no control over the local
strain, and the system exhibits local, macroscopic  strain inhomogeneity~\cite{zhang2010coarse}, as shown in Fig.~\ref{fig:apparatus}(f). In particular,
during the course of a strain experiment, it tends to develop a shear
band.
\IncludeFigureApparatus
For both types of devices, the initial state is prepared by placing
the particles within the boundaries of the container, with the
particles lying on the corresponding base. In general, there are
residual forces acting between the grains after the placement of the
particles. We remove these by gently tapping or massaging the grains
by a small amount. Thus, the initial state is force-free for the
experiments described here. Necessarily, that means that they lie below
what we call $\phi_J$; for larger $\phi$, all static states are jammed
at non-zero pressure. In a typical experiment, the system is subject
to small quasi-static strain steps, up to some maximum shear
strain. { For protocol I, the strain step is $ 0.27\% $ and for protocol II, it is $ 0.3\% $.} After each small step, the system is allowed to come to mechanical equilibrium,
after which we obtain images that characterize the state of the
system. We carry out different types of shear experiments that include
shear to some maximum strain, and cyclic shear, where the system is 1)
sheared from an initial state to a maximum shear strain, then returned
to its initial boundary configuration, and 2) then subject to repeats
of this protocol.

The particles are illuminated from below by a circularly polarized
uniform light source. They are also illuminated from above by a low
intensity UV light source. The systems are imaged from above by a
camera which obtains three different images of each state, following a
given small strain step. One image is acquired with a crossed circular
polarizer in front of the camera, a second image is acquired without
that crossed polarizer, and a third is acquired with only the UV light
source. The first type of image gives the photoelastic response of the
system, the second gives the location and boundaries of the particles,
and the third shows the orientation of small bars that have been drawn
of each particle with UV sensitive ink to track rotation.

{
Since protocol I creates homogeneous shear-jammed states without shear bands,  they are a much better candidate for analyzing the nature of shear-jamming and shear-jammed states.  Analysis of inhomogeneous states with shear bands such as those created by protocol II are more difficult to characterize.  Therefore, in this paper, we have focused most of our analysis on the states created by protocol I.  However,  we have compared some features of the shear jamming transition from the two protocols in section IV.C to demonstrate the difference between the patterns in height space in states with and without shear bands.}

\subsection{Analysis Methods}
\subsubsection{Construction of Force Tiles}
Topologically, the network of force tiles is the graph dual to the real-space contact network {(a graph with grains as nodes and force-bearing contacts between grains as edges)}. This duality is unique in 2D.  Hence, given the topology of the real-space network (RSN) and the information about the minimal cycles of the RSN, the topology of the force tile network (FTN) can be easily constructed in the form of an adjacency matrix. The height points are then constructed from the topological information as well as the information about the forces that act through the contacts. 
The  difficulty in designing a numerical algorithm to construct the topology of a dual graph is associated with the construction of the minimum cycle basis (MCB). 
 We adapted an existing algorithm~\cite{MCB_Algo} for constructing  MCBs to generate the adjacency matrix required to construct the FTNs. As discussed in section I.A, height vectors  are represented by the vertices of the FTN, whereas the edges represent the forces acting through the contacts. The choice of the height origin and, therefore,  the choice of  a particular vertex as the origin of the FTN, is arbitrary.   However, once this is chosen, any force and torque balanced grain configuration gives rise to a unique set of vertices. A metric (Euclidean) for the FTN is defined  by assigning a  scalar weight (magnitude of the force) and a direction (direction of the force through a contact) to each edge in a FTN. The object that encodes all of the information for constructing a FTN is, therefore, a vector-weighted adjacency matrix.  Using this adjacency matrix, the FTN can be constructed iteratively. This iterative construction process is analogous to constructing a lattice from the primitive vectors. An important difference is that, in the case of a two-dimensional lattice, there are only two primitive vectors, whereas due to the disordered nature of the FTN, there are as many primitive vectors as there are contacts, which necessitates an iterative process. The entire construction process is summarized below and in Fig.~\ref{fig:FT_Algo}: 
\begin{itemize}
\item Obtain RSN from microscopic information.
\item Obtain the topology of the dual graph of the RSN. This is also the topology of the FTN and is encoded in an adjacency matrix with entries that are zero and unity.
\item Replace the non-zero entries of the adjacency matrix by two numbers that correspond to the components of the force vector  acting through that particular contact. 
\item Start from a trial configuration of the height space vertices, and construct them iteratively using the adjacency matrix with weights as described above.  The iteration ends when the final difference between two heights (vertices of the FTN)  is the same (within a small numerical tolerance) as the vector weight of the entry in the adjacency matrix corresponding to the edge connecting these two vertices. 
\end{itemize} 
\IncludeFigureFTAlgo

From an algorithmic point of view the first step, obtaining the RSN, is the easiest. The contacts and force at the granular level are easily obtained from either simulation or experimental data. Some amount of post-processing is required to obtain the FTN for experimental data, since Newton's third law may not always be rigorously satisfied due to experimental limitations. However, this shortcoming can be accommodated within the framework of the algorithm to construct FTN. Since we already have the contact topology for the FTN, it ensures force balance by default.  All the forces acting on a grain are forced to form a closed polygon. Hence, we can enforce Newton's third law on a contact by taking the average of $ \vec f_{ij} $, and $ \vec f_{ji} $. This allows us to construct a FTN, which is force-balanced and where Newton's third law is satisfied. The forces constructed this way are accurate within the experimental errors.  Data obtained from simulations would be cleaner (margin of error much smaller), and would satisfy both force balance and Newton's third law. Hence, no post-processing would be required to implement the scheme.   In this paper, we have analyzed only experimental data obtained from the shearing experiments described in the previous section.


\section{Results}  

\subsection{Shear jamming transition}

 The mechanical properties of shear jammed materials are different from those of an elastic solid. Since by design ($\phi < \phi_J$) the systems that we are studying have zero {inter-granular} forces  at zero shear stress, unlike an elastic solid, this zero-stress state in not a well-defined reference state. These zero-shear states cannot resist any mechanical perturbation and behave like a fluid: these are the unjammed states~\cite{corey_simul}.  As the external shear stress is increased, the granular material transitions from being unjammed to a  fragile solid  to a solid that can resist shear reversals~\cite{BiNature2011}.  This is quite unlike an elastic solid, which deforms reversibly and ultimately undergoes plastic failure  at large enough shear stress:  external shear does not strengthen an elastic solid.   The origin of this strengthening in granular materials can be ultimately traced back to the lack of cohesive forces, which leads to a differential mechanical response of force chains in the dilational and compressive directions.  Our objective in this paper is to understand the implication of these microscopic processes on the collective behavior of grains.   The primary question that we are focusing on  is what type of correlations develop during the shear-jamming process that leads to the formation of jammed states that can resist further shearing.

Rigidity of crystalline solids is associated with the emergence of order and broken translational symmetry  in the averaged density field  of the constituent particles. This broken symmetry exists even in an  amorphous solid:  if one measures the thermal average of the density field, it is uniform for a liquid ($ \langle \rho\rangle = \rho_0 $), but nonuniform ($ \langle\rho \rangle \ne \rho_0 $) for an amorphous solid.   The shear-jamming transition occurs at zero temperature, and according to previous analysis~\cite{BiNature2011,Cates1998}  the organization that leads to the emergence of rigidity is primarily in the space of forces.  As illustrated in Fig. \ref{fig:grain_position}, apart from the global affine deformation, the positions of the grains remain virtually unchanged throughout the shear-jamming process, and any measure of the density field can at most reveal subtle changes.  


\IncludeFigureGrainPosition 


The situation is remarkably different  for FTNs  (Fig. \ref{fig:FT_SJ})  at different strain steps: there is a clear evolution of the both the global shapes and the distribution of shapes of individual tiles.  At the beginning of the shear protocol,  FTNs are not well defined since very few grains form a force carrying network. As the shear strain is increased, FTNs emerge but both their global shape and local structure changes from one strain step to another.  In this regime, the global shapes are also more needle like (one-dimensional).  As the shear strain is increased further, the FTNs acquire a well-defined two dimensional structure, and deform uniformly as the strain is increased, with only small changes in the local structures.   {The question we address is whether the patterns in RSNs and FTNs become persistent during the shear jamming protocols.}


\IncludeFigureForceTiling


\subsubsection{{Characterizing the order in SJ states}} As mentioned earlier, characterizing the ``order'' in any type of amorphous solid is a non-trivial problem.  One of the ideas that has been applied extensively in glassy systems is that of measuring the persistence of patterns through an overlap matrix.  In lattice models such as spin glasses, for example~\cite{RevModPhys.58.765}, similarity between two different replicas, $ \alpha $ and $ \beta $ at a given temperature,  is measured by the overlap matrix $ Q^{\alpha,\beta} $: 
\begin{eqnarray}
Q^{\alpha,\beta}_i  = \langle s^\alpha_i s^\beta_i\rangle, \quad \alpha \neq \beta,
\end{eqnarray}
where $ s_{\alpha}^i $ is the $ i^{th} $ spin of the replica $ \alpha $. The angular bracket denotes thermal average. Each element of the overlap matrix varies between $ 0 $ and $ 1 $. In the high temperature phase,  the average overlap for a system of $ N $ spins, $ \bar{Q} = \frac 1 N \sum_{i} Q^{\alpha,\beta}_i $ , is zero, whereas in the spin-glass phase $ \bar{Q} $ is nonzero. 
One can construct similar overlap matrices for continuum systems, and this has been done in the context of  molecular glasses~\cite{dasgupta1991there,dasgupta2000free}, where one analyzes the overlap of coarse-grained density fields corresponding to different free energy minima.  {  The overlap measure of order is similar to measuring autocorrelation functions.  The autocorrelation function is an overlap of configurations at different times and thus explicitly measures the persistence of patterns.  A non-vanishing correlation in the limit of infinite time is equivalent to a finite overlap~\cite{RevModPhys.58.765}.   the Edwards-Anderson order parameter in spin glasses is the large-time limit of the spin autocorrelation function~\cite{edwards1975theory}.

In this section we analyze whether an overlap matrix measure can be used to identify the SJ transition.
We first define coarse-grained density fields corresponding to the point pattern of vertices in the FTNs and the position of grains in RSN.  An overlap matrix between different strain steps can then be constructed from these density fields in a manner analogous to glassy systems.}
In analyzing the overlap of FTNs, one of the complications that we need to address  is that in the shear jamming experiments, the number of height vertices increases with the shear strain and the shape and the area of the box enclosing the height vertices changes.  For RSNs, the box only changes shape while the area and the number of points (grains) remains fixed.    We, therefore, need to supplement the usual coarse graining prescription by constructing a grid that distorts affinely with the box,  and by normalizing the coarse-grained density field appropriately.   With these modifications, the overlap matrix is defined as: 
\begin{eqnarray}
Q^{\alpha,\beta} = d^{\alpha,\beta}/\sqrt{d^{\alpha,\alpha}d^{\beta,\beta}}
\label{eq:overlap},
\end{eqnarray} 
where $ d^{\alpha,\beta} = \frac 1 N\sum_{m = 1}^{N}\rho^\alpha_m \rho^\beta_m  $, and  $ \rho^\alpha_m $ is the value of the coarse-grained density field of the $ \alpha^{th} $ point pattern (corresponding to the $ \alpha^{th} $ strain step)  at the $ m $th  grid point.



\paragraph*{Overlap as a measure of rigidity:} If one point pattern can be obtained from the other solely through a series of affine transformations, the overlap as defined by Eq. \ref{eq:overlap}  between those two patterns should be unity.  As  two point patterns, modulo an affine transformation, deviate away from each other , the overlap also decreases towards the minimum value of 0.  For example,  a linearly  elastic deformation of a crystalline solid  from the zero-shear reference state is a completely affine process; the position of the atoms, which form the point pattern, in one state can be obtained from the other through a series of scaling, rotation and translation. Thus, the overlap between two such states will be unity. On the other hand, if a liquid is sheared, the molecular displacements are less restricted, and it is unlikely that the point pattern of the sheared state is related to the  unsheared state through any affine transformation.  {This analogy suggests that shear-rigidity can be related to the overlap between configurations at different shear strains. } 

Associating shear rigidity with the properties of an overlap matrix is particularly advantageous for the granular systems that we are interested in, since we do not need to define a zero stress reference state. We measure the overlap of two states at  any two strain steps, $\alpha$ and $\beta$, and obtain one matrix for every experimental run: the zero strain step is not treated in any special way.


\IncludeFigureOLComparison


\subsubsection{Emergence of rigidity}
We calculate  the overlap of the coarse grained density fields from RSNs and FTNs at different strains, $\gamma$. For all the analysis in this paper, we have used a $ 30\times30 $ grid to measure the coarse-grained density field. To accommodate  the change in the sizes of the point pattern, we scale all the point patterns to a $ 1\times1 $ box, from which we calculate the density field by counting the number of points in each of the boxes created by the grid. The density field depends on the coarse-graining size (grid size in this case) and, as the coarse-graining size is increased (number of grid points is decreased), the density field becomes more uniform across different grids. Consequently, the overlap between two such pattern increases. However, there is a range of coarse-graining sizes over which the pattern is robust.    

 As seen from  Fig. \ref{fig:OL_Comp}a, the overlap matrix constructed from the RSNs is relatively structureless and {the} majority of the elements are $ \sim 1 $. There is an initial small range of strains for which the overlap decays to zero quickly with incremental strain.  Beyond $\gamma \approx 10\%$, however, the overlap between any two strain steps is of order unity. The overlap matrices vary from run to run but these broad features remain unchanged, indicating  that the RSN patterns are persistent over the whole strain history, for the full range of $\phi$ values below $\phi_J$ for which shear-jamming is observed.  The implication of this result is that from the perspective of a coarse-grained density field, the assembly of grains at these packing fractions is in an amorphous ``solid'' state.  However, these amorphous solids cannot resist shear deformations and, therefore, are not rigid.  They acquire shear-rigidity only beyond a threshold of applied shear.   Remarkably, the overlap matrix obtained from FTNs bears a signature of this rigidity transition.

The overlap matrix of the FTNs  shown in Fig. \ref{fig:OL_Comp}b exhibits a non-trivial pattern.  The most remarkable feature of this pattern is an increase in the range of strains{, $\delta\gamma = \beta - \alpha$,} over which the overlap remains significantly above zero. It is clear from Fig. \ref{fig:OL_Comp}b that this range increases monotonically with the initial strain $\alpha$.  The implication is that there is a clear signature of the shear-jamming transition  in FTNs:  the persistence of the height pattern.  We can define an order parameter by thresholding the entries of the overlap matrix : $q^{\alpha,\beta}= 1~ {\rm if}  ~ Q^{\alpha,\beta}  \ge 0.5 $, and $q^{\alpha,\beta}= 0~ {\rm if} ~  Q^{\alpha,\beta}  < 0.5$.   An ``order parameter'' $\epsilon^*(\gamma)$ can then be defined as the range $\delta \gamma$ over which $q^{\alpha,\beta} = 1$.   {Although different from the traditional definition of an order parameter,  we call $\epsilon^*(\gamma)$ {an} order parameter for shear-jamming since it is zero in the fragile regime and non-zero in the shear-jammed regime as seen from Fig.~\ref{fig:epsilon}.   Moreover, $\epsilon^*(\gamma)$ offers a quantitative measure of the rigidity of the SJ states:  states that have a larger value of $\epsilon^*(\gamma)$ can sustain larger shear strains.   As we will discuss in the next sections, the behavior of $\epsilon^*(\gamma)$ is different for the two protocols.   For {protocol II,} which creates states with shear bands, $\epsilon^*(\gamma)$ is non-monotonic indicating that the SJ states can become weaker under shearing.   Experiments show that the SJ states created by {protocol II} can undergo failure through avalanches, and this feature is captured by $\epsilon^*(\gamma)$.   As we will  also show in the next section,  $\epsilon^*(\gamma)$ exhibits a scaling behavior with packing fraction for the homogeneous SJ states created by {protocol I}.}

\subsection{Role of the packing fraction}
We observe shear jamming for a range of packing fractions below $ \phi_J $. In protocol I, we observe shear jamming for $ \phi \in [0.74, 0.8247]$ and in protocol II, we observe shear jamming for $ \phi \in [0.78, 0.825] $. In the following paragraphs we investigate the role of the packing fractions in shear jamming transition by analyzing experimental data from protocol I.   { One of the main results of this analysis is the deduction of a scaling form for the non-rattler fraction and the order parameter, which indicates that the structure of the shear-jammed states is controlled by a scaling combination $\gamma/\gamma(\phi)$ with $\gamma(\phi) \rightarrow 0$ as $\phi \rightarrow \phi_J$.}


    
%

We have investigated the dependence of the SJ transition on $\phi$ by comparing the stress anisotropy, $\tau/P$, the non-rattler fraction, $ f_{NR} $, the order parameter, $ \epsilon^* $,  and the FTN overlap matrices for five different packing fractions: 0.8269, 0.8163, 0.8036, 0.7863 and 0.7728. We show the strain dependence of  $ f_{NR} (\gamma, \phi) $ and $ \epsilon^*(\gamma, \phi) $ in Fig.~\ref{fig:epsilon} and of $ (\tau/P) (\gamma,\phi)$ in Fig.~\ref{fig:JRComparison}(b) for all five packing fractions. The data shown for each packing fraction is averaged over five different runs. Furthermore, we show the overlap matrix for  $ \phi_1 = 0.8163 $ and $ \phi_2 = 0.8036 $ in Figs.~\ref{fig:JRComparison}(c-d). We choose these two values of $\phi$ because all three regimes  of  shear jamming -- unjammed, fragile and jammed -- are clearly captured in these packing fractions. For larger packing fractions, the system very quickly transitions into the jammed state, so the behavior of the system in the fragile state is difficult to decipher. On the other hand, lower packing fractions do not reach the shear jammed state within the range of $\gamma$ explored. 

\IncludeFigureEpsilonStar

Figs.~\ref{fig:JRComparison}(b)-(d) show that the strain required to reach the jammed state is different for different packing fractions. More precisely, the strain required to reach the jammed state increases as the packing fraction decreases. {Based on the experimental observations, it has been postulated that the strain required to reach the jammed state diverges at a packing fraction $ \phi_s $~\cite{BiNature2011,Bi2012Thesis, ren2013reynolds}.} The exact value of  $ \phi_s $ is not known. However, experimental measurements estimate it to be around $ 0.74 $ for protocol I~\cite{Ren2013Thesis}. At this packing fraction, the system does not jam even at the largest experimental strain ($ \sim 70 \% $).   Simulations of frictionless disks investigating particle scale reversibility find that there is a maximum strain that increases with decreasing packing fraction, which demarcates regions of point reversible and loop reversible dynamics~\cite{Schreck:2013fk}.  The similarity between this behavior of the maximal strain and the strain required to reach shear jamming is intriguing and should be explored further in simulations of frictional grains.

The upper limit of shear-jamming packing fractions is set by $\phi_J$ at which the strain required to create a jammed packing goes to zero.   
The FTN overlap matrices show this trend, which can be made quantitative by analyzing the order parameter $\epsilon^* (\gamma,\phi)$. A similar analysis can be performed for  $ f_{NR} $. As shown in  Fig.~\ref{fig:epsilon}(a) and (d), both $ f_{NR} $ and $ \epsilon^* $ depend strongly on the packing fraction.  As the packing fraction increases towards $ \phi_J $, the system transitions to the shear jammed state at a smaller strain. 
The strain dependence is particularly remarkable for $ \epsilon^* $ with the transition becoming sharper as the packing fraction is increased.

{The above observation suggests that there is a packing fraction dependent strain $ \gamma_0(\phi) $ characterizing the shear jamming transition. Rescaling $ \gamma $ by $ \gamma_0 (\phi) = (1- \phi/\phi_J)^{1.6}$  leads to a good scaling collapse of  $ \epsilon^*  (\gamma,\phi)$ and of $ f_{NR} (\gamma,\phi)$ except in the very small  $\gamma$ regime (Figs.~\ref{fig:epsilon} b,e).   
The insight that we gain from this scaling analysis is that although the SJ transition is seemingly controlled by two parameters, $\phi$ and $\gamma$,  it is only the scaling combination $\gamma/\gamma_{0}(\phi)$ that is relevant. A scaling form for  $ f_{NR} (\gamma,\phi)$ offers a natural explanation for the  observed  increase in Reynolds pressure with packing fraction ~\cite{ren2013reynolds}.  

Reynolds pressure is a consequence of the shear dilatancy of the frictional grains. When frictional grains are sheared under constant pressure, the granular packing dilates. In contrast, in our experiments, the area of the granular packing is held constant which frustrates the dilation of the packing leading to an increase  in the pressure of the granular packing. This pressure is observed to increase as $ R(\phi)\gamma^2 $~\cite{ren2013reynolds}, where the Reynolds coefficient,  $ R(\phi) $,   diverges as $ (1-\phi/\phi_J)^{-3.3\pm0.1}$. The pressure, therefore, has a scaling form: $ P(\phi,\gamma) \sim (\gamma/\gamma_0^p (\phi))^2 $ with $ \gamma_0^p  (\phi) \sim (1-\phi/\phi_J)^{1.65 \pm 0.05}$.  The similarity  between the observed scaling exponents of $\gamma_0 (\phi) $ and $\gamma_0^p (\phi)$   is striking and suggests that the pressure increase is a direct consequence of the increase in $ f_{NR} $ and that the divergence of $R(\phi)$ at $\phi_J$ is related to the scaling of $\gamma(\phi)$. }



{  The scaling form shown in Figs.~\ref{fig:epsilon} (b) and (e) fails   to collapse the data for $f_{NR}$ at low values of strain primarily because the the value of $ f_{NR} $ at $ \gamma = 0 $ strongly depends on the packing fraction as shown in Fig.~\ref{fig:epsilon}(a).  To account for this, we have attempted a different scaling collapse by defining a scaled variable $ y_{sc}(\gamma)\equiv \frac{y(\gamma) - y(0)}{y_{max} - y(0)} $.  Setting $ f_{max} = 1 $, we find  $ f_{sc} \sim \gamma/\gamma_1 (\phi)$, where $ \gamma_1 (\phi) = (1-\phi/\phi_J)^{1.2} $ (Fig.~\ref{fig:epsilon}c). $ \epsilon^* $ exhibits a similar scaling law  if $ \epsilon_{max} $ is defined by the saturation values of the $ \epsilon^* $ (Fig.~\ref{fig:epsilon}f).  This form of  scaling  of $f_{NR}$ had earlier been predicted  by a phenomenological theory of the the shear-jamming transition~\cite{Sarkar2015Framework}. }

\IncludeFigureJRComparison



\subsection{Protocol dependence of the SJ transition}

We observed shear jamming in both protocol I and {II}.   Even though the basic phenomenology -- saturation of $ f_{NR} $, percolation of the strong force network at the jamming transition -- is the same in both experiments, the difference in the protocols affect the macroscopic variables significantly. To recall, in protocol I, which we have discussed so far, shear strain is applied homogeneously across the sample. On the other hand, in protocol II, shear strain is applied from the boundary and that leads to strain inhomogeneity in the system. This strain inhomogeneity leads to formation of shear bands, which affect the stability of the jammed structure. {  In particular, these jammed states can fail under shear whereas those created by protocol I do not {generally} show macroscopic failure}.
In the following paragraphs, { we provide some comparisons of the different measures of shear jamming in the two protocols at $ \phi = 0.8036 $ ($ \phi_J \approx 0.84$ for both protocols) in order to illustrate the effects of shear banding and failure on the order parameter and the statistics of the FTNs.}


\IncludeFigureJRJZFT


 We applied simple shear in protocol I and pure shear in protocol II. The former does not preserve the compressive and the extensile direction over the course of the experiment, while the latter does. As a result,  the FTNs  change both shape and orientation during the shearing process of protocol I. In contrast,  they  just change shape in protocol II. Since the system is sheared from the boundary in protocol II, the forces {can be} higher near  one boundary compared to the other. This non-uniformity of forces is captured by the FTNs. As shown in Fig.~\ref{fig:JRJZ_FT}, the force tiles at the bottom left corner of the FTN are much larger than the rest of the tiling. The shapes of the individual tiles are also quite different. While the polygonal tiles generated by  protocol I are more or less regular in shape, those generated by  II can be quite irregular. This observation can be made more quantitative by measuring the distribution of the asphericity (see Appendix for definition) of the polygons (Fig.~\ref{fig:Asphericity}). The mean asphericity of the tiles in protocol II is always higher than protocol I.   
 \IncludeFigureAsphericity
 
 \IncludeFigureJRJZThree
 The larger dispersion in tile shapes and sizes in the FTNs from protocol II indicates a broader distribution of contact forces and grain-level stresses.   We see this feature in the distribution of the magnitude of the contact forces.    Even in the SJ state, the distribution generated by protocol II has an exponential tail (Fig. \ref{fig:ForceDistro}(c)), which is characteristic of a marginally jammed state~\cite{corey_simul}, whereas  the distribution generated by protocol I becomes narrower and develops a well defined peak as states become shear-jammed (Fig. \ref{fig:ForceDistro}(a-b))
%

These broad distributions suggest that the SJ states created via protocol II are less rigid than those created by protocol I.  Comparison of the FTN overlap matrices from the two protocols support this view.  Fig. \ref{fig:JRJZ_1} (c-d) illustrate the differences between the persistence of FTN patterns created by the two different protocols.  The FTNs generated by protocol II are much less  persistent, and the structure of the overlaps do not really evolve with $\gamma$.  The order parameter, $\epsilon^*$ shown in Fig.~\ref{fig:JRJZ_1} (e) provides a quantitative measure of  the difference in rigidity between SJ states created by these two protocols.  { In particular, $\epsilon^*$ exhibits a non-monotonic behavior in protocol II reflecting the shear-induced failure of jammed states created by this protocol.   This result shows that shear bands influence the persistence of patterns in FTNs and that their effect can be measured by the order parameter $\epsilon^*$.  The scaling of $f_{NR}$ and $\epsilon^*$ in protocol I, shown in Fig. \ref {fig:epsilon},  is also not observed for protocol II.}

\IncludeFigureJRJZOne


\subsubsection{Percolation of force chains and force tilings}

\IncludeFigurePercolation
The original analysis of shear jamming~\cite{BiNature2011} was based on  the structure of the force network formed as a function of applied strain. It is known from earlier studies that force networks in jammed packings of dry grains can be separated into a strong network and a complementary weak network~\cite{Radjai}.   The shear-induced solidification framework~\cite{Cates1998} was also based on characterization of the force network.  Here, we present a comparison of the force-network analysis with the force-tiling analysis of the shear-jamming process in protocol II.   The tiling representation explicitly includes only non-rattlers and it has only the information about the topology of the fabric of granular contacts.  The ``spatial'' information in the tiling space represents the forces.  In contrast, the percolation analysis explicitly incorporates the real-space fabric~\cite{BiNature2011} and the percolation analysis is based on a thresholding of forces.

To perform the percolation analysis, we first define the collection of grains that have at least one contact force $\vec{f}_{ij}$ where  $f_{ij} > \langle f_{ij}\rangle$~\cite{BiNature2011}:  a \emph{strong force carrier}. The strong force carrier grains form multiple clusters in the system with a distribution of sizes. In turn, we define the largest such cluster the \emph{strong force cluster}. Grains with contact forces $f_{ij}\le  \langle f_{ij}\rangle$
are part of a complementary force network or weak force network.   As shown earlier~\cite{BiNature2011,Radjai}, the results of the percolation analysis are robust as long as the threshold is chosen to be $\ge 0.8 \langle f_{ij}\rangle$.

In Fig.~\ref{fig:clusters}, we show the size of the cluster as function of strain in a typical strain cycle. Starting from an unjammed state,  as shown in Fig.~\ref{fig:clusters}, the strong force cluster undergoes two sequential percolation transitions as a function of the strain.  Initially, the system is unjammed and  the strong force cluster does not percolate in either the compressive direction or the extension direction, i.e., $\xi_x<L_x$ and  $\xi_y<L_y$. The reason that an unjammed state with a a non-percolating force network can exist in this system is because there are contacts between the grains and the substrate, which allows force and torque balance to be satisfied for a subset of grains.
At intermediate strain values, the strong force cluster percolates in the compressive direction but not transverse to it, i.e., $\xi_y=L_y$ and $\xi_x<L_x$.   We call these states fragile~\cite{Cates1998,BiNature2011}. At higher strain values, the strong force cluster percolates in both directions $\xi_y=L_y$ and $\xi_x=L_x$ and we call these states shear-jammed.

The force tilings corresponding to these same strain steps are shown in the bottom panel of Fig.~\ref{fig:clusters}, and the overlap measure, $\epsilon^*$ is shown  in Fig.~\ref{fig:JRJZ_1}(e).   
{ 	
	The persistence of tiling patterns, measured by the overlap function provides a direct measure of the resistance to shear of the jammed state.   If the force network in a jammed state rearranges under shear, then the overlap function is small.  The magnitude of $ \epsilon^* $ measures {the} range of shear strain {over which} the jammed state maintains its rigidity.  The percolation analysis, on the other hand,  tells us whether a particular jammed state has a force network that has percolated in one or two directions (in 2D).  It is an observation that when the network has percolated in 2D, it {resists} shear and is shear-jammed.  However, the percolation measure does not tell us how much shear this state can resist.  This is precisely what the overlap function does, as illustrated by the non-monotonicity of $\epsilon^*$ is shown  in Fig.~\ref{fig:JRJZ_1}(e).  To summarize, the percolation analysis is a binary measure that does not provide information about how strong a shear-jammed state is, i.e., how much shear it can resist.  The overlap function and $  \epsilon^* $ provides this information and, therefore, they are not only complementary to the percolation analysis but they provide additional useful information about the shear-jammed states.   
}

\subsection{Convexity of tiles and the rigidity of the jammed packing} 

The discussions in the previous two subsections demonstrate (figures~\ref{fig:JRComparison} and \ref{fig:JRJZ_1}) that the overlap matrix convincingly captures the rigidity of the shear jammed states and the lack of the rigidity of the fragile states. We now inquire into the origin of the rigidity in the shear jammed states. 

At the end of section II, we hypothesized that the persistence of patterns in FTNs has its origin in effective interactions between height vertices created by the condition of convexity of tiles. Although we do not have a complete understanding of the nature of these interactions and the correlations they generate, analysis of experiments show a strong correlation between the persistence of the pattern and the statistics of convex tiles. In Fig.~\ref{fig:JRJZ_NNC} (a) and (b), we show the variation of the anisotropic stress ($ \tau $) and the fraction of non-convex polygons  relative to the maximum ($ N_{NC} $),  as a function of the applied shear strain for protocol I. The evolution of  these  two variables with strain are remarkably similar. As seen from  Figs.~\ref{fig:JRJZ_NNC}(a) and (b), the position of the peak in $ N_{NC} (\gamma)$ coincides with that of the  peak in  $ \tau (\gamma)$.  Remarkably,  the position of these peaks coincide with the value of $\gamma$ at which the order parameter $ \epsilon^* $ reaches its maximum value (Fig.~\ref{fig:epsilon}). We can, therefore,  conclude that $N_{NC}$ starts decreasing with $\gamma$ in the SJ phase, where the order parameter $\epsilon^*$ is non-zero.  These observations, however,  do not demonstrate  a causal relationship between the behavior of $N_{NC}$ and the development of the order parameter.  The shear-jamming process of protocol II does not lead to a well-developed order parameter, and,  therefore, to a well-defined SJ state (Fig.~\ref{fig:JRJZ_1}). Consistent with our hypothesis, we find that $ N_{NC} $ does not decrease (Fig.~\ref{fig:JRJZ_NNC}(c)) with increasing shear strain in protocol II. 
           
\IncludeFigureJRJZTwo

\section{Discussion}
Shear-jamming of frictional grains is phenomenologically very different from the traditionally-studied jamming of frictionless grains. Frictionless grains undergo a  density-induced jamming transition as the packing fraction is increased to a characteristic value, $\phi_J$, which can depend on the protocol.  Frictional grains exhibit a much richer jamming phenomenology.   In these systems, imposed shear strain can induce a jamming transition over a range of packing fractions below $\phi_J$.   The primary focus of our work has been to construct a rigorous theoretical framework for describing these shear-jamming transitions.

In this paper, we have applied the dual-space formalism based on the FTN representation of granular assemblies in mechanical equilibrium to characterize shear-jamming transitions observed under two different experimental protocols.   Our analysis clearly identifies signatures of the jamming transition through the properties of an overlap matrix and a resulting scalar order parameter.   The overlap matrix and the order parameter are sensitive to the nature of the jammed states created by different protocols.  A particularly striking result of our analysis is that the strength of the order is weaker in {shear-jammed} states with shear bands as demonstrated in Fig. ~\ref{fig:JRJZ_1}(e)).   Moreover, we find that for the homogeneous states created by protocol I,  the effect of the packing fraction can be captured by rescaling the shear strain marking the onset of shear-jamming.  These observations indicate that the dual-space formalism is the natural  representation for characterizing shear jamming, which is difficult to detect in position space.  The FTN representation is equally applicable to density-driven jamming in both frictional and frictionless grains.  Hence, this representation  can be the common thread which unifies  the study of frictionless and frictional jamming.

The origin of the emergence of ``order'' in the FTNs is the set of constraints of  local mechanical equilibrium for dry grains.   The necessary condition for persistent order is the geometrical constraint of convexity on the shape of force tiles formed by connecting the heights corresponding to a single grain. This geometrical constraint is a consequence of two inequalities: positivity of the normal forces, and the  static equilibrium restriction on the range of the tangential  forces.  Persistent order develops as more and more force bearing contacts are introduced into a grain packing, which translates to an increase in the number of height vertices.  The process is thus reminiscent of density-driven solidification, albeit in a space that refers to forces and not positions of grains.    We will explore this analogy more carefully in the near future.

The FTN representation  provides a description of elastic and plastic behavior of assemblies of dry grains by referring only to their stress state specified through $\vec F_x$ and $\vec F_y$.  This stress-only description avoids any reference to the concepts of strain and energy, which are difficult to define unambiguously in assemblies of dry grains~\cite{Cates1998}.


Our analysis has been restricted to 2D.  The tiling picture does not extend to 3D.  An analog of the height fields does exist in 3D~\cite{Henkes,Degiuli}, and a completely parallel structure can be constructed through Delaunay triangulation of the grain network in real-space~\cite{Degiuli}.   
It is, therefore, plausible that the general concept of order in height space extends to 3D, and  \cite{Degiuli} provides a mathematical framework for developing and testing a theory of rigidity in 3D.

\begin{acknowledgments}
We acknowledge extended discussions with Joshua Dijksman, Dong Wang, Jonathan Bares, Thibault Bertrand, Mark Shattuck, and Corey O'Hern. SS would like to thank Mitch Mailman for sharing his data, which was used to generate Fig.~\ref{fig:FT_Algo}. 

BC, SS and DB were supported by NSF-DMR 0905880 $\&$ 1409093. RPB was supported by NSF DMR-1206351, NSF-DMS-1248071, NASA NNX15AD38G. SS, RPB, and BC were supported by the W. M.  Keck foundation, and NSF-PHY11-25915. JZ acknowledges the Chinese 1000-Plan (C) fellowship, Shanghai Pujiang Program (13PJ1405300), and National Natural Science Foundation of China (11474196).
\end{acknowledgments}

\appendix*

\section{}
\subsection*{Appendix I:  Experimental Methods}

In both protocols, we tested dozens of packing densities, and up to five runs at each density, within a range $ \phi_S < \phi < \phi_J $, where$  \phi_J \simeq 0.84 $ is the isotropic jamming point in 2D, and $  \phi_S \simeq 0.75 $ is the minimum density for shear jamming~\cite{zhang2010jamming,BiNature2011}. This density range ensures that the system starts from a completely stress-free state, while making sure shear jamming will develop with increasing strain. At the beginning of each experiment, we prepared a disordered packing by manually rearranging the particles. We also gently tapped or pushed particles to remove all forces in the system in order to start the experiment from a stress-free state.

{All experiments described here used particles, e.g. disks, made of photoelastic material to obtain contact forces between particles. The use of photoelasticity to obtain inter-particle forces was first described in Majmudar and Behringer~\cite{majmudar2005contact}, in more detail in the Ph.D. theses by Majmudar~\cite{Majmudar2006Thesis} and Ren~\cite{Ren2013Thesis}, and elsewhere \cite{majmudar2007jamming,zhang2010statistical,zhang2010coarse,BiNature2011,ren2013reynolds}. In order to understand the basic principle of the force-finding procedure, we note that if a particle is subject to a known set of contact forces (which we imagine are roughly point forces), then the stresses within the particle are known. For instance, for a finite number of point-like contact forces acting on a disk, the stresses within the disk are given in terms of a closed form solution that involve the vector contact forces. Circularly polarized light of intensity $I_o$ and wavelength $ \lambda $ that first traverses a disk of thickness $ T $ along a ray that is normal to the plane of the disk and then a crossed circular polarizer, has an intensity $I = I_o \sin^2[C T \pi (\sigma_2 - \sigma_1)/\lambda]$. Here $ C $  is the stress optic coefficient of the photoelastic material and the $\sigma_i$ are the planar principle stresses at each point in the disk. Determining the interparticle contact forces involves solving a non-linear inverse problem that seeks the contact forces which yield the observed photoelastic response within the disk. This inverse procedure is implemented on each disk independently, where the algorithm expressly incorporates force and torque balance on each particle.}

{\textbf{ Protocol I}} provides simple shear to collections of photoelastic disks. Fig.4 (a-b) shows a schematic of this shear protocol. The unique feature of this apparatus is its capability to provide shear strain that is spatially very uniform, modulo small local fluctuations. The base of the cell consists of 50 parallel narrow slats, each with width 12.7mm. Each slat is individually tied to, and co-moves with, the long opposing walls of the cell.  In its undeformed state, the apparatus has interior dimensions of 30cm by 60cm. During shear, the slats move uniformly and act to carry the particles sitting on them, providing an affine shear background to the system. The particles are 8.0mm and 6.4mm in diameter, with 1:3.3 large-to-small number ratio. Fig.4 (c) demonstrates that the resulting particle movement is largely uniform with small and uncorrelated fluctuations, and that shear bands or other macroscopic inhomogeneities are absent.

{\textbf{Protocol II }} is sketched in Fig. 4 (d-e). The boundaries of this system are controlled so as to produce pure shear, consisting of compression in one direction, and dilation in the orthogonal direction. The particles rest on a base consisting of a smooth Plexiglas sheet that is powder lubricated to reduce the friction between the particles and the Plexiglas sheet, and are confined by the boundaries, whose positions are controlled by a pair of stepper motors (not shown). The particles are bidisperse, 7.4mm and 8.6mm in diameter, with 1:1.5 large-to-small number ratio in this apparatus. Fig.4 (d-e) represents protocol (II), and the resulting particle movement of this protocol (Fig.4 (f)) shows that the system exhibits local, macroscopic strain inhomogeneity (shear banding). The upper boundary is fixed in the frame of the base. The three other walls move as indicated in part (d). That is, in order for the system to evolve from (c) to (d), the three lower boundaries move in the indicated directions relative to the base and top wall. The side walls are maintained in a rectangular geometry by sliders. This apparatus allows a deformation of the boundaries in a continuous range of rectangular geometries. However, for the experiments described here, the area of the interior, which contains the particles, is held fixed. Hence the strains correspond to pure shear. In the device, the strain is strictly applied at the boundaries, as is typical of most granular strain devices. Consequently, there is no control over the local strain, and the system exhibits local, macroscopic strain inhomogeneity [34], as shown in Fig. 4(f). In particular, during the course of a strain experiment, a shear band tends to develop. 

In the first protocol, we carried out multiple runs for 11 different packing fractions from $\phi \simeq 0.69$ to $\phi \simeq 0.82$. In each case, the initial state was prepared stress-free. In the second protocol, we prepared the stress-free and homogeneous initial states in a total of 100 different packing fractions equally spaced between $\phi_{min} =0.792$ and {$\phi_{max} = 0.850$}. To eliminate any potential correlations between run to run, the particle configurations of each run were freshly prepared. 

 In both types of experiments, a horizontal layer of frictional photo-elastic disks (particles) was quasi-statically sheared in small equal strain increments. The incremental strain step is {0.27\% in protocol (I) and 0.3\%} in protocol (II), chosen as the minimum step that could be accurately and reliably achieved with each apparatus. In each protocol, the system consisted of roughly 1,000 particles that were bi-disperse in size to ensure disordered packing. The particle size selection was slightly different for the two protocols because of the apparatus dimensions and practical considerations. 
 
After each increment, we took three photos using different lighting conditions in order to record the position, orientation, and photo-elastic force response of all particles. Also, for both types of devices, the initial state was prepared by placing the particles within the boundaries of the container, with the particles lying on the corresponding base. In general, there were residual forces acting between the grains after the placement of the particles. We removed these by gently tapping or massaging the grains by a small amount. Thus, the initial state was force-free for the experiments described here. 

The mechanical and statistical analysis of shear jamming dynamics in these experiments have been reported elsewhere, based on physical measures like the fraction of force-bearing particles, pressure, $f_{NR}$, the shear stress, and force statistics ~\cite{zhang2010jamming,BiNature2011}. In the main sections of this paper,  we have imported the particle position and contact forces into our force tiling algorithm and carried out our theoretical analysis. 

\subsection*{Appendix II: Correspondence between FTN and RSN}
The construction of the FTN helps us evaluate the partition function of a granular system from the geometry of the force tiles. So, we expect to obtain the principal stress eigenvalues or, equivalently, the pressure and the stress anisotropy from the geometry of the FTN. This correspondence becomes particularly simple in a 2D granular system with periodic boundary condition (PBC), which we describe here. We claim that this result is far more general and works quite well for experimental systems where PBC is not employed. 

As we have seen in II.A.2, the extensive stress tensor (also called the force-moment tensor) of the 2D system under PBC can be written as: 
\bea
\hat{\Sigma} &=& \hat{L} \times \hat{F} \\
&=&\begin{pmatrix}
L_x	& 0\\
0 & L_y
\end{pmatrix}\times \begin{pmatrix}
F_{xx} & F_{xy}\\
F_{yx} & F_{yy}
\end{pmatrix}\\
&=&\begin{pmatrix}
L_xF_{xx} & L_xF_{xy}\\
L_yF_{yx} & L_yF_{yy}
\end{pmatrix},
\eea 
where the $ \hat{F} $ tensor codifies the global shape of the FTN. The trace of the force-moment tensor $ \hat \Sigma $ is the ``global pressure" $ P $ or the isotropic component of the stress and it is an invariant of the matrix. Hence, $ 2P = \Sigma_1 +\Sigma_2= L_xF_{xx}+L_yF_{yy} $, where $ \Sigma_{1,2} $ are the eigenvalues of $ \hat \Sigma $. On the other hand, the anisotropic stress $ \tau = \abs{\Sigma_1 - \Sigma_2} = \sqrt{P^2 - L_xL_y(F_{xx}F_{yy}- F_{xy}F_{yx})}$. 

We start the geometric interpretation of these results by noting that $(F_{xx}F_{yy}- F_{xy}F_{yx})  $ is the area of the parallelogram bounded by $ \vec{F_x} $ and $ \vec{F_y} $. Additionally, if we assume that $ L_x = L_y = L $, and $ F_{xy} = F_{yx} = 0 $, FTN and RSN becomes uncoupled and then the geometric correspondence becomes more transparent. Under this assumption, $ P = L(F_{xx}+F_{yy})/2 $, which is one quarter of the perimeter of the rectangle bounded by the $ \vec{F_{x,y}} $ vectors. Similarly, $ \tau = L*\sqrt((\mbox{Perimeter/4})^2 -\mbox{Area} ) $. In the more general case, when  FTN  and  RSN  are coupled, the perimeter of the FTN isn't exactly equal to $ P $, but provides a good estimate as illustrated in Fig.~\ref{fig:Stress_Geometry_Corr}(a) and (c). The area of the tiles on the other hand provides an excellent estimate of the determinant of the stress tensor Fig.~\ref{fig:Stress_Geometry_Corr}(b) and (d). The pressure, which is the average of the stress eigenvalues and the determinant which is the product of the stress eigenvalues, can be used to estimate the stress anisotropy $ \tau $, which measures the difference in the eigenvalues.     
  
\paragraph*{Shape Anisotropy} The anisotropy of the stress state may be approximated by measuring the shape anisotropy of the tiles. This is achieved through calculating the gyration tensor of the tiles. The gyration tensor is defined as $ S_{mn} = \frac 1 N \sum_{i= 1}^N r_m^{(i)}r_n^{(i)} $. The shape anisotropy is then measured through \textit{asphericity} $ \kappa^2 = 2\frac{\lambda_1^4+\lambda_2^4}{\left(\lambda_1^2+\lambda_2^2\right)^2} - 1 $, where $ \lambda_1 $ and $ \lambda_2 $ are the eigenvalues of the two dimensional gyration tensor. For a regular polygon, the relative shape anisotropy is zero, and for a line it's exactly one. 



\IncludeFigureStressGeometryRelation


\bibliography{Master_Ref}
\end{document}